\input amstex
\documentstyle{amsppt}
\magnification=\magstephalf

 \addto\tenpoint{\baselineskip 15pt
  \abovedisplayskip18pt plus4.5pt minus9pt
  \belowdisplayskip\abovedisplayskip
  \abovedisplayshortskip0pt plus4.5pt
  \belowdisplayshortskip10.5pt plus4.5pt minus6pt}\tenpoint
\pagewidth{6.5truein} \pageheight{8.9truein}
\subheadskip\bigskipamount
\belowheadskip\bigskipamount
\aboveheadskip=3\bigskipamount
\catcode`\@=11
\def\output@{\shipout\vbox{%
 \ifrunheads@ \makeheadline \pagebody
       \else \pagebody \fi \makefootline 
 }%
 \advancepageno \ifnum\outputpenalty>-\@MM\else\dosupereject\fi}
\outer\def\subhead#1\endsubhead{\par\penaltyandskip@{-100}\subheadskip
  \noindent{\subheadfont@\ignorespaces#1\unskip\endgraf}\removelastskip
  \nobreak\medskip\noindent}
\def\endremark{\par\revert@envir\endremark\vskip\postdemoskip}
\outer\def\enddocument{\par
  \add@missing\endRefs
  \add@missing\endroster \add@missing\endproclaim
  \add@missing\enddefinition
  \add@missing\enddemo \add@missing\endremark \add@missing\endexample
 \ifmonograph@ 
 \else
 \vfill
 \nobreak
 \thetranslator@
 \count@\z@ \loop\ifnum\count@<\addresscount@\advance\count@\@ne
 \csname address\number\count@\endcsname
 \csname email\number\count@\endcsname
 \repeat
\fi
 \supereject\end}
\catcode`\@=\active
\PSAMSFonts
\CenteredTagsOnSplits
\NoBlackBoxes
\nologo
\def\today{\ifcase\month\or
 January\or February\or March\or April\or May\or June\or
 July\or August\or September\or October\or November\or December\fi
 \space\number\day, \number\year}
\define\({\left(}
\define\){\right)}
\define\Ahat{{\hat A}}
\define\Aut{\operatorname{Aut}}
\define\CC{{\Bbb C}}
\define\CP{{\Bbb C\Bbb P}}

\define\Hom{\operatorname{Hom}}
\define\Map{\operatorname{Map}}

\define\RR{{\Bbb R}}
\define\SS{\Bbb S}
\define\Spin{\operatorname{Spin}}

\define\Tr{\operatorname{Tr}}
\define\ZZ{{\Bbb Z}}
\define\[{\left[}
\define\]{\right]}
\define\ch{\operatorname{ch}}
\define\chiup{\raise.5ex\hbox{$\chi$}}
\define\cir{S^1}

\define\exertag #1#2{#2\ #1}

\define\index{\operatorname{index}}

\define\inv{^{-1}}
\define\mstrut{^{\vphantom{1*\prime y}}}
\define\protag#1 #2{#2\ #1}
\define\rank{\operatorname{rank}}
\define\res#1{\negmedspace\bigm|_{#1}}
\define\temsquare{\raise3.5pt\hbox{\boxed{ }}}

\define\theprotag#1 #2{#2~#1}

\define\xca#1{\removelastskip\medskip\noindent{\smc%
#1\unskip.}\enspace\ignorespaces }

\define\zmod#1{\ZZ/#1\ZZ}

\define\zt{\zmod2}

\define\rem#1{\marginalstar\begingroup\bf[{\eightpoint\smc{#1}}]\endgroup}
\def\strutdepth{\dp\strutbox} 
\def\marginalstar{\strut\vadjust{\kern-\strutdepth\specialstar}} 
\def\specialstar{\vtop to \strutdepth{ 
    \baselineskip\strutdepth 
    \vss\llap{$\bold{\Rightarrow}$ }\null}}

\NoRunningHeads 


\loadeusm
\input xy \xyoption{all} \redefine\cir{S^1}

\define\Alpha{A}
\define\BSpin{B\text{Spin}}
\define\Det{\operatorname{Det}}
\define\Dir#1{D\mstrut _{\X/S}(#1)}
\define\Dirac{D\hskip-.65em /} 
\define\GG{G\times G}
\define\Lie{\operatorname{Lie}}
\define\MSpin{M\text{Spin}}
\define\RZ{\RR/\ZZ}
\define\Sign{\operatorname{Sign}}
\define\TT{\Bbb{T}}
\define\Tu{\Theta ^{\text{univ}}}
\define\UU{\Bbb{U}}
\define\VV{\Bbb{V}}
\define\WV{\Omega ^{V}}
\define\WW{\Bbb{W}}
\define\WXS{\Omega ^{\X/S}}
\define\X{\eusm{X}}
\define\bG{\overline{\Gamma }}
\define\bQ{\overline{Q}}
\define\bU{\overline{U}}
\define\bV{\overline{V}}
\define\bg{\bar{\gamma }}
\define\boson#1{\Cal{B}_{#1}}
\define\cE{\check E}
\define\cH{\check{H}}
\define\cUe{\check{U}_e}
\define\ca{\check\alpha }
\define\cb{\check\beta }
\define\ce{\check{e}}
\define\cm{\check\mu }
\define\cn{\check\nu }
\define\conn#1#2{\eusm{A}_{#1}^{(#2)}}
\define\cv{_{\text{cv}}}
\define\drx#1{\dot\rho (\xi _{#1})}
\define\eE{\eta \mstrut _{\sE}}
\define\field#1{\Cal{F}_{#1}}
\define\filt{\operatorname{filt}}
\define\hol{\operatorname{hol}}
\define\id{\operatorname{id}}
\define\point{\text{point}}
\define\sE{\Cal E}
\define\sG{\Cal{G}}
\define\sH{\Cal{H}}
\define\sL{\Cal{L}}
\define\sP{\eusm{P}}
\define\sQ{\Cal{Q}}
\define\tG{\tilde{\Gamma }}
\define\tW{\widetilde{W}}
\define\tboson#1{\widetilde{\Cal{B}}_{#1}}
\define\tfield#1{\widetilde{\Cal{F}}_{#1}}
\define\tg{\tilde{\gamma }}
\define\tl{\tilde{\lambda }}
\define\tpi{2\pi i}
\define\triv{\bold{1}} 
\define\wm{\omega _{\cm}} 
\define\wn{\omega _{\cn}} 
\define\zo{[0,1]}

\refstyle{A}
\widestnumber\key{SSSSS}   
\document

 \pretitle{$$\boxed{\boxed{\text{REVISED VERSION}}}$$\par\vskip 3pc}

	\topmatter
 \title\nofrills Pions and Generalized Cohomology \endtitle
 \author Daniel S. Freed  \endauthor
 \thanks The work of D.S.F. is supported in part by NSF grant DMS-0603964.  I
also thank the Aspen Center for Physics for warm hospitality during the final
stages of this work.\endthanks
 \affil Department of Mathematics \\ University of Texas at Austin\endaffil 
 \address Department of Mathematics, University of Texas, 1 University
Station C1200, Austin, TX 78712-0257\endaddress 
 \email dafr\@math.utexas.edu \endemail
 \date May 31, 2007\enddate
 \dedicatory In memory of Raoul Bott \enddedicatory
 \abstract  
 The Wess-Zumino-Witten term was first introduced in the low energy $\sigma
$-model which describes pions, the Goldstone bosons for the broken flavor
symmetry in quantum chromodynamics.  We introduce a new definition of this
term in arbitrary gravitational backgrounds.  It matches several features of
the fundamental gauge theory, including the presence of fermionic states and
the anomaly of the flavor symmetry.  To achieve this matching we use a
certain generalized differential cohomology theory.  We also prove a formula
for the determinant line bundle of special families of Dirac operators on
4-manifolds in terms of this cohomology theory.  One consequence is that
there are no global anomalies in the Standard Model (in arbitrary
gravitational backgrounds).
 \endabstract
	\endtopmatter

\document

Quantum chromodynamics has a global symmetry group~$G\times G$, where
$G=SU_{N_f}$ in the theory with $N_f$~flavors of massless quarks.  This is
presumed broken to the diagonal, with the homogeneous space $(G\times G)/ G$
parameterizing the vacua.  The low energy dynamics of the Goldstone
bosons---the pions---is modeled by a nonlinear $\sigma $-model with
target~$(G\times G)/G$; see~\cite{We} for an account.  There is a topological
term in the $\sigma $-model action, first introduced by Wess and
Zumino~\cite{WZ} and later elaborated by Witten~\cite{W1}; see also~\cite{N}.
We propose a new, geometric definition of this term (\theprotag{4.1}
{Definition}).  Our motivation is to reproduce certain features of the high
energy gauge theory in the low energy effective theory.  By working in the
Euclidean theory formulated on arbitrary Riemannian spin 4-manifolds---in
other words, by studying the theory in an arbitrary Euclidean gravitational
background---we are able to probe more than can be seen in flat
spacetime.\footnote{Indeed, previous treatments use spherical
compactifications of spacetime, and because the Hurewicz map $\pi _5G\to
H_5G$ is not an isomorphism certain features were missed.}  Specifically:
 \roster
 \item"(i)" a spin structure is required to define this new WZW term, just as
a spin structure is necessary in the high energy theory to define spinor
fields;
 \item"(ii)" canonical quantization naturally gives a $\zt$-graded Hilbert
space, matching the presence of both bosonic and fermionic states in~QCD, and
the statistics formula~\thetag{4.10} which follows directly from our
definition is correct;
 \item"(iii)" our definition works for $N_f=2$; and
 \item"(iv)" there is a natural gauged version of the theory whose anomaly
matches that of gauged~QCD.
 \endroster
  This last property is an example of 't Hooft anomaly matching~\cite{'tH},
and it serves to fix the coefficient of the WZW~term, as in~\cite{W1}.  We
emphasize the second point, that fermionic states appear in a theory with
only bosonic fields.  This phenomenon also occurs in theories with self-dual
fields~\cite{FMS1}, but via a different mechanism.  We verify properties
(i)--(iv) in~\S{4}.

To capture all of these features we use a {\it generalized cohomology
theory\/} where previously ordinary integral cohomology was used.
Generalized cohomology theories, especially the various forms of $K$-theory,
appear in high energy theoretical physics in connection with
anomalies---through the Atiyah-Singer index theorem---and as a home for the
Dirac quantization of Ramond-Ramond charges in superstring theory.  The
theory we encounter here also appears when defining three-dimensional
Chern-Simons theory on spin manifolds~\cite{J}.  It has also appeared
recently~\cite{FHT}, \cite{AS} in connection with twisted $K$-theory, though
that connection plays no role here.  This special cohomology theory, which we
simply denote~`$E$' and elucidate in~\S{1}, has exactly two nontrivial
homotopy groups, so is a simple twisted product of two ordinary cohomology
theories.\footnote{Although most familiar are geometric cohomology theories,
such as $K$-theories and cobordism theories, generalized cohomology theories
are like boutiques: abundant and specialized.  The theory~$E$ is a minimal
choice for this problem, but presumably not a unique one.}  It is a natural
home for a characteristic class of complex vector bundles which plays the
role of~$c_3/2$, half the third Chern class (see \theprotag{1.9}
{Proposition}).  This allows us to prove in~\S{2} that the isomorphism class
of the determinant line bundle of special families of Dirac operators on spin
4-manifolds is computed as an integral in~$E$.  There are analogous
specialized index theorems in dimensions one~\cite{FW,(5.22)} and
two~\cite{F,\S5} in terms of ordinary cohomology.  As the dimension grows,
the denominators in the Riemann-Roch formula grow, and so the index recedes
further from ordinary cohomology.  The small denominator of~2 in our problem
permits the formula in terms of the cohomology theory~$E$.

It is critical in our applications that these topological ideas be promoted
to theorems in generalized {\it differential\/} cohomology~\cite{HS}.  Thus
\theprotag{2.2} {Theorem} is a formula in differential $E$-cohomology for the
determinant line bundle with its covariant derivative.  Ordinary differential
cohomology, also known as smooth Deligne cohomology or the theory of
Cheeger-Simons differential characters, was first used by Gaw\c
edzki~\cite{G} to express terms of Wess-Zumino-Witten type.  In~\S{4} we use
differential $E$-theory to define the Wess-Zumino-Witten term in the
effective $\sigma $-model for pions.  The main point is that there is a class
~$\nu \in E^5(G)$ which is in a precise sense half the generator of~$H^5(G)$,
as we prove in \theprotag{1.9} {Proposition}.  The {\it gauged\/} WZW term,
as defined in \theprotag{4.14} {Definition}, depends on a certain {\it
transgression\/} which occurs when attempting to extend~$\nu $ to a
$(\GG)$-equivariant class.  (The connection between the gauged WZW~term and
transgression was made in ordinary cohomology in~\cite{W4,Appendix}.)  The
class~$\nu $ does not so extend: the obstruction is the anomaly, which is the
transgression of~$\nu $, and the transgressing `$E$-cochain' is used to
define the gauged WZW term.  This whole discussion must be carried out in the
{\it differential\/} theory,\footnote{See \theprotag{5.12} {Definition},
which is an extension of ideas in~\cite{HS}.}  so requires the construction
of a transgressing differential form~\thetag{5.16}.\footnote{See the recent
preprint~\cite{Jo} which treats a much generalized version of this
transgression problem.}

The index formula proved in \theprotag{2.2} {Theorem} applies to compute
anomalies in four-dimensional gauge theories.  Here we mean the full anomaly,
including global anomalies.  Indeed, in \theprotag{3.4} {Example} we prove
that the Standard Model has no global anomalies.  We apply the index formula
to compute the anomaly in gauged~QCD (\theprotag{3.7} {Proposition}).  Also,
the characteristic class~$\mu $ leads to the definition (\theprotag{3.10}
{Definition}) of an {\it anomaly-free subgroup\/} which captures both local
and global anomalies.

There are other models of QCD with different flavor symmetry groups and so
different homogeneous spaces in the low energy effective $\sigma $-model;
see~\cite{MMN}, \cite{DZ} for example.  Presumably there is a similar story
for the WZW term in these cases as well, but we leave it for others to
investigate.

Raoul had a tremendous influence on me, both mathematically and personally.
His passion for mathematics and music was infectious, his passion for life
inspiring.  He was a great teacher in every sense of the word.  I'd like to
think that this mix of geometry, topology, and physics---with a touch of
transgression thrown in---would be to his taste.

It is a pleasure to thank Jacques Distler, Mike Hopkins, Greg Moore, Sonia
Paban, and Lorenzo Sadun for valuable discussions and comments.

 \head
 \S{1} $E$-Theory
 \endhead
 \comment
 lasteqno 1@ 18
 \endcomment

In topology a cohomology theory is specified by a {\it spectrum\/}, which is
a sequence~$\{E_n\}_{n\in \ZZ}$ of pointed topological spaces and maps
$\Sigma E_n\to E_{n+1}$ such that the adjoint maps $E_n\to \Omega E_{n+1}$
are homeomorphisms.  Here $\Sigma $ denotes suspension\footnote{The
suspension of a pointed space~$X$ is the pointed space 
  $$ \Sigma X = [0,1]\times X\bigm/ \{0\}\times X \;\cup\; \{1\}\times X
     \;\cup\; [0,1]\times \{x_0\},  $$
where $x_0 \in X$ is the basepoint.} and $\Omega $~the based loop space.  The
cohomology of a space~$X$ is then the abelian group
  $$ E^n(X) = [X,E_n] \tag{1.1} $$
of homotopy classes of maps into the spectrum.  The Eilenberg-MacLane
spectrum~$HA$ attached to an abelian group~$A$ is characterized by 
  $$ \pi _qHA_n = \cases A ,&q=n;\\0,&\text{otherwise}.\endcases  $$

The spectrum~$E$ of interest in this paper, which for lack of good
alternatives we simply notate~`$E$', has two nontrivial homotopy groups.  The
$n^{\text{th}}$~space~$E_n$ fits into the fibration\footnote{This is written
without reference to particular spaces in the spectra as $H\ZZ \to E \to
\Sigma ^{-2}H\zt$.}
  $$ \xymatrix@1{H\ZZ_n \ar[r]^-i &E_n \ar[r]^-j &H\zt_{n-2}} \tag{1.2} $$
whose classifying map $H\zt_{n-2}\to H\ZZ_{n+1}$ is the stable cohomology
operation~$\beta \circ Sq^2$, the integer Bockstein composed with the second
Steenrod square.  The fibration~\thetag{1.2} leads, for any space~$X$, to a
long exact sequence
  $$ \xymatrix@1{\cdots \ar[r] &H^n(X) \ar[r]^-i &E^n(X) \ar[r] ^-j
     &H^{n-2}(X;\zt) \ar[r]^-{\beta \circ Sq^2} & H^{n+1}(X) \ar[r] &\cdots
     } \tag{1.3} $$
Since multiplication by~2 on $H\zt_{n-2}$ is homotopically trivial,
multiplication by~2 on~$E_n$ lands in the image of~$i$, so defines the
map~$k$ in the diagram 
  $$ \xymatrix{& H\ZZ_n \ar[d]^-i \\
                E_n\ar@{-->}[ur]^-k\ar[r]^-2 & E_n} \tag{1.4} $$
Notice that any class in the image of~$i\circ k$ is divisible by~2 and that
$k\circ i$~is multiplication by~2: for any space~$X$ the composition 
  $$ \xymatrix{& H^n(X) \ar[r]^-i & E^n(X) \ar[r] ^-k & H^n(X)} \tag{1.5} $$
is multiplication by~2.
 
In low degrees we have for any space~$X$ 
  $$ \alignedat2
      E^0(X) &\cong H^0(X)&&\cong [X,\ZZ], \\
      E^1(X) &\cong H^1(X)&&\cong [X,\RR/\ZZ].\endaligned  $$
There is a short exact sequence
  $$ 0\longrightarrow H^2(X) \longrightarrow E^2(X)\longrightarrow
     H^0(X;\zt)\longrightarrow 0,  $$
and we can interpret~$E^2(X)$ as the group of isomorphism classes of
$\zt$-graded complex line bundles on~$X$.  Similarly, there is a short exact
sequence
  $$ 0\longrightarrow H^3(X) \longrightarrow E^3(X)\longrightarrow
     H^1(X;\zt)\longrightarrow 0,  $$
and we can interpret~$E^3(X)$ as the group of isomorphism classes of
$\zt$-graded gerbes.

$E$-theory is oriented for spin manifolds; see \theprotag{4.4} {Proposition}
for details.  Thus given a map $f\:X\to Y$ whose stable relative normal
bundle $f^*TY-TX$ carries a spin structure, there is an {\it umker\/} or
pushforward map
  $$ f_*\:E^{\bullet}(X)\longrightarrow E^{\bullet - n}(Y), \tag{1.6} $$
where $n=\dim X - \dim Y$.
 
The differential theory~$\cE$ associated to~$E$ is defined on smooth
manifolds~$M$; see~\cite{HS,\S4} and also~\cite{FMS2,\S2} for an expository
introduction to some aspects of differential cohomology (in general, not
specifically for this theory~$E$) which is geared to physicists.  We will not
repeat the definitions here (except in \theprotag{5.10} {Definition}) but do
note the following exact sequence~\cite{HS,(4.57)}.  Define
  $$ A^n(M) = \left\{ (\lambda ,\omega )\in E^n(M)\times \Omega
     ^n_{\text{closed}}(M): [\omega ]_{\text{dR}} = \frac 12k(\lambda )_{\RR}
     \right\}, \tag{1.7} $$
where $k(\lambda )_{\RR}\in H^n(M;\RR)$ is the image of $k(\lambda )\in
H^n(M)$ under the natural map $H^n(M)\to H^n(M;\RR)$ and $[\omega
]_{\text{dR}}\in H^n(M;\RR)$ is the de Rham cohomology class of the closed
differential form~$\omega $.  Then the sequence
  $$ 0 \longrightarrow E^{n-1}(M)\otimes \RR/\ZZ \longrightarrow
     \cE^n(M)\longrightarrow A^n(M)\longrightarrow 0 \tag{1.8} $$
is exact.  The differential form associated to a differential cohomology
class is termed its {\it curvature\/}.  The differential $E$-groups in low
degrees also have geometric interpretations: $\cE^1(M)$~is the topological
abelian group of maps $M\to\RR/\ZZ$ and $\cE^2(M)$~is the topological abelian
group of isomorphism classes of $\zt$-graded hermitian line bundles with
unitary covariant derivative.  Integration in~$\cE$ is defined over a fiber
bundle~$X\to Y$ whose stable normal (or tangent) bundle is given a spin
structure.
      
After these generalities we turn to the construction of the characteristic
class~$\mu $.  We use the standard notions~$w_n,p_n,c_n$ for Stiefel-Whitney,
Pontrjagin, and Chern classes.  We also use the notation $h_3,h_5$ for
generators of~$H^{3}(SU_N),H^5(SU_N)$.  Only the mod~2 reduction of the
former enters; the sign of the latter, defined for~$N\ge3$, is fixed
in~\thetag{1.15} below.

        \proclaim{\protag{1.9} {Proposition}} \newline
 \indent{\rm(i)}\ There is an isomorphism $E^4(BSO)\cong \ZZ$ with
generator~$\lambda $ satisfying 
  $$ k(\lambda )=p_1,\qquad j(\lambda )\equiv w_2\pmod2.  $$
The pullback of~$\lambda $ to~$\BSpin$ is~$i(\tl)$ for a class~$\tl\in
H^4(\BSpin)$ with~$2\tl=p_1$.\newline
 \indent{\rm(ii)}\ For~$N\ge3$ there is an isomorphism $E^6(BSU_N)\cong \ZZ$ with
generator~$\mu $ satisfying 
  $$ k(\mu ) = c_3,\qquad j(\mu )\equiv c_2\pmod2.  $$
Also, $E^6(BSU_2)\cong \zt$ and the generator~$\mu $ satisfies $j(\mu )\equiv
c_2\pmod2$. \newline 
 \indent{\rm(iii)}\ For all~$N\ge1$ there is an isomorphism $E^6(BSp_N)\cong
\zt$.  The generator is the pullback of~$\mu $ under $BSp_N\to
BSU_{2N}$.\newline
 \indent{\rm(iv)}\ For $N\ge3$ there is an isomorphism $E^5(SU_N)\cong \ZZ$
with generator~$\nu $ satisfying
  $$ k(\nu ) = h_5,\qquad j(\nu )\equiv h_3\pmod2. \tag{1.10} $$
The class~$\nu $ is invariant under left and right translation by~$SU_N$.
Also, $E^5(SU_2)\cong \zt$ and the generator~$\nu $ satisfies $j(\nu )\equiv
h_3\pmod2$.\newline
 \indent{\rm(v)}\ The characteristic class~$\mu $ obeys a Whitney sum
formula: if $V_1,V_2$~are complex vector bundles with trivialized
determinants, then 
  $$ \mu (V_1\oplus V_2) = \mu (V_1)+\mu (V_2). \tag{1.11} $$
Also, $\mu (\overline{V})=-\mu (V)$.
        \endproclaim

\flushpar
 Note that $c_2\pmod2$ is the pullback of~$w_4$ under $BSU_N\to BSO$.  The
class~$\lambda $ plays the role of `$\frac 12p_1$' and the class~$\mu $ plays
the role of~`$\frac 12c_3$'.   

        \demo{Proof}
 For~(i) we construct a map $BSO\to E_4$ by attaching cells of dimension $\ge
6$ to~$BSO$ to kill~$\pi _qBSO,\,q\ge 5$; the space so constructed is~$E_4$.
This gives an element~$\lambda \in E^4(BSO)$ for which $j(\lambda )=w_2$
and~$k(\lambda )=p_1$.  Now use~\thetag{1.3} and the map $\BSpin\to BSO$ to
obtain a commutative diagram in which the rows are exact:
  $$ \xymatrix{ H^1(BSO;\zt) \ar[d] \ar[r]^-0 &H^4(BSO)\ar[d]^-a
     \ar[r]^-{i_1} &E^4(BSO) \ar[d]\ar[r] &H^2(BSO;\zt) \ar[d]\ar[r]^-{\beta
     \circ Sq^2} & H^5(BSO) \ar[d]\\
      H^1(\BSpin;\zt) \ar[r]^-0 & H^4(\BSpin) \ar[r]^-{i_2} &E^4(\BSpin)
     \ar[r] &H^2(\BSpin;\zt) \ar[r]\ar[r]^-{\beta \circ Sq^2} &H^5(\BSpin)} $$
In the top row $H^4(BSO)\cong \ZZ$ with generator~$p_1$; $H^2(BSO;\zt)\cong
\zt$ with generator~$w_2$; and $\beta w_2^2=0$ since $w_2^2\equiv p_1\pmod2$.
Thus $E^4(BSO)\cong \ZZ$ or~$\ZZ\times \zt$, but the existence of~$\lambda
\in E^4(BSO)$ with $k(\lambda )=p_1$ rules out the latter,
using~\thetag{1.5}.  In the second row of the diagram $H^2(\BSpin;\zt)=0$,
from which $i_2$~is an isomorphism.  Also, $H^4(\BSpin)\cong \ZZ$ and we can
choose a generator~$\tl$ such that~$a(p_1)=2\tl$.
 
For~(ii) we note that the space $BSU_N$ has a CW~presentation $BSU_N\sim
S^4\cup e_6\cup \cdots$, valid for~$N\ge3$.  The attaching map $\partial
e_6\to S^4$ is~$\eta $, the double suspension of the Hopf map $S^3\to S^2$;
it represents the nontrivial element of~$\pi _5S^4\cong \zt$.  (That the
attaching map is nontrivial follows from the relation~$Sq^2c_2\equiv c_3\pmod
2$, for example.)  The long exact sequence in $E$-cohomology attached to the
cofibration
  $$ S^4\cup_\eta e_6 \longrightarrow BSU_N\longrightarrow
     (BSU_N,S^4\cup_\eta e_6) \tag{1.12} $$
shows that $E^6(S^4\cup_\eta e_6)\cong E^6(BSU_N)$, since the CW~presentation
of the quotient starts with an 8-cell.  By the suspension isomorphism 
  $$ E^6(S^4\cup_\eta e_6)\cong E^4(S^2\cup_\eta e_4)\cong
     \widetilde{E}^4(\CP^2). \tag{1.13} $$
In~\thetag{1.13} the attaching map~$\eta \:\partial e^4\to S^2$ {\it is\/}
the Hopf map, and $\widetilde{E}^4$~is the reduced $E$-cohomology, which in
this case is isomorphic to the unreduced $E$-cohomology.  Then the exact
sequence
  $$ 0 \longrightarrow H^4(\CP^2)\longrightarrow  E^4(\CP^2)\longrightarrow
     H^2(\CP^2;\zt)\longrightarrow 0  $$
shows that $E^4(\CP^2)\cong \ZZ$ or~$\ZZ\times \zt$.  But the underlying real
2-plane bundle~$V$ to the hyperplane line bundle over~$\CP^2$ has
$k\bigl(\lambda (V) \bigr)=p_1V$ the generator of~$H^4(\CP^2)$.  It follows
that $E^4(\CP^2)\cong \ZZ$. 
 
The argument for~(iv) is similar.  There is a CW~presentation $SU_N\sim
S^3\cup e_5 \cup e_7 \cup\cdots$; the attaching map $\partial e_5\to S^3$
is~$\eta $, the suspension of the Hopf map, since $\pi _5SU_N\cong \ZZ$.
Analogous arguments to~\thetag{1.12} and~\thetag{1.13} show $E^5(SU_N)\cong
E^5(S^3\cup_\eta e_5)\cong \widetilde{E}^4(\CP^2)\cong \ZZ$.  The invariance
of~$\nu $ under translation is immediate from the homotopy invariance of
cohomology, since $SU_N$~is connected.
 
Let $p\:BSU_{N_1}\times BSU_{N_2}\to BSU_{N_1+N_2}$ be the direct sum map.
Then \thetag{1.11}~is equivalent to the equation $p^*(\mu _{N_1+N_2})=\mu
_{N_1}+\mu _{N_2}$.  The difference between the two sides is a class
$c\in E^6(BSU_{N_1}\times BSU_{N_2})$ which satisfies $j(c)=0$, by the
Whitney formula for Chern classes, and so from~\thetag{1.3}, $c=i(b)$ for some
$b\in H^6(BSU_{N_1}\times BSU_{N_2})$.  But the Whitney formula for Chern
classes also implies that $k(c)=0$, and so $b$~is torsion of order~2
by~\thetag{1.4}.  Since $H^6(BSU_{N_1}\times BSU_{N_2})$ is torsionfree, we
deduce that~$c=0$.  A similar argument proves that $\mu $~changes sign under
conjugation.

To prove~(iii) we use~\thetag{1.3} and the vanishing of~$H^6(BSp_N)$
and~$H^7(BSp_N)$ to show that $E^6(BSp_N)@>j>> H^4(BSp_N;\zt)$ is an
isomorphism; the latter group is cyclic of order~2.  Furthermore, the
commutative square 
  $$ \xymatrix{E^6(BSU_{2N}) \ar[d]\ar[r]^-j &H^4(BSU_{2N};\zt) \ar[d]\\
                E^6(BSp_N)\ar[r]^-j &H^4(BSp_N;\zt)}  $$
shows that $\mu \in E^6(BSU_{2N})$ maps to the generator of~$E^6(BSp_N)$,
since $j(\mu )=w_4$ maps to the generator of~$H^4(BSp_N;\zt)$.

The statements for~$SU_2$ and~$BSU_2=BSp_1$ follow directly from the long
exact sequence~\thetag{1.3}.
        \enddemo

Next, we promote~$\mu ,\nu $ to differential classes, and for that we
need\footnote{We refer to~\cite{HS,\S3.3} for a more elegant treatment of
differential characteristic classes which uses the Weyl algebra of
polynomials on the Lie algebra in place of differential forms on the
classifying space.} a smooth model of the classifying spaces.  Let $\sH$~be
an infinite dimensional separable complex Hilbert space which carries a
quaternionic structure.  Let $ESU_N$~be the Stiefel manifold of isometries
$\CC^N\to \sH$ and $BSU_N$~the quotient by the natural right $SU_N$~action.
Also, let $ESp_N$~be the submanifold of~$ESU_{2N}$ of isometries $\CC^{2N}\to
\sH$ which preserve the quaternionic structure and $BSp_N$ the quotient by
the natural right $Sp_N$~action.  Note the natural map $BSp_N\to BSU_{2N}$
and its lift to the universal principal bundles.  The universal bundle
$ESU_N\to BSU_N$ carries a connection~$\Tu$ defined by the orthogonal
complements to the orbits; let $\Omega $~denote its curvature.  Now it
follows easily from~\thetag{1.3} that $E^5(BSU_N)=0$ and $E^4(SU_N)=0$, so
\thetag{1.8}~implies that to promote~$\mu ,\nu $ to differential classes
$\cm\in \cE^6(BSU_N)$, $\cn\in \cE^5(SU_N)$ we have only to specify closed
differential forms ~$\wm\in \Omega ^6(BSU_N)$ and~$\wn\in \Omega ^5(SU_N)$.
Let $\theta \in \Omega ^1(SU_N;\frak{s}\frak{u}_N)$ denote the Maurer-Cartan
form, often written~$\theta =g\inv dg$.  Then 
  $$ \align
      \wm &= \frac{-i}{48\pi ^3}\,\Tr\Omega ^3, \tag{1.14}\\
      \wn &= \frac{-i}{480\pi ^3}\,\Tr\theta ^5. \tag{1.15}\endalign $$
The differential form~\thetag{1.15} is bi-invariant, i.e., invariant under
both left and right translation in~$SU_N$.  Since $\nu $~is also
bi-invariant, it follows that so too is~$\cn$.

Let $P\to M$ be a principal $SU_N$-bundle with connection~$\Theta $ over a
smooth manifold~$M$ and $V\to M$ the associated rank~$N$ hermitian vector
bundle with covariant derivative.  We define a {\it differential
characteristic class\/}~$\cm(V)\in \cE^6(M)$.  For simplicity\footnote{We
could instead use a classifying map for the bundle and a Chern-Simons form
which measures the difference between~$\gamma ^*\Tu$ and~$\Theta $;
cf.~\cite{HS,\S3.3}.} we use the existence of an $SU_N$-equivariant map
$\gamma \:P\to ESU_N$ which pulls the universal connection~$\Tu$ back
to~$\Theta $.  In fact, the space of such classifying maps is contractible
and nonempty~\cite{DHZ}.  A classifying map induces $\bg\:M\to BSU_N$, and we
set $\cm(V)=\bg^*(\cm)$.  The ``curvature'' of~$\cm(V)$ is the 6-form
  $$ \omega _{\cm(V)} = \frac{-i}{48\pi ^3}\Tr(\WV)^3, \tag{1.16} $$
where $\WV$~is the curvature of~$V$. 

        \proclaim{\protag{1.17} {Lemma}}
 $\bg^*(\cm)$~is independent of the classifying map~$\bg$. 
        \endproclaim

        \demo{Proof}
 Since any two classifying maps are homotopic, and the 6-form curvature is
independent of~$\bg$, the image of~$\bg^*(\cm)$ in~$A^6(M)$
(see~\thetag{1.8}) is independent of~$\bg$.  Let $\Gamma \:[0,1]\times P\to
ESU_N$ be a homotopy of classifying maps of~$\Theta $, and $\bG\:[0,1]\times
M\to BSU_N$ the quotient homotopy.  Then $\bG^*_1(\cm) - \bG_0^*(\cm)$~is in
the image of~$E^{n-1}(M)\otimes \RR/\ZZ$, so can be detected by a map of a
closed spin $(n-1)$-manifold $f\:W^{n-1}\to M$.  By Stokes' theorem
  $$ \int_{W}\,f^*\bG_1^*(\cm) - f^*\bG_0^*(\cm) \;=\; \int_{[0,1]\times
     W}\bigl(\operatorname{id}_{[0,1]}\times f \bigr)^*\bG^*\wm. \tag{1.18} $$
But $\bG^*\wm\in \Omega ^6\bigl([0,1]\times M \bigr)$ is the pullback
of~\thetag{1.16} via projection to~$M$, whence \thetag{1.18}~vanishes. 
        \enddemo

 \head
 \S{2} Determinant Line Bundles on 4-Manifolds
 \endhead
 \comment
 lasteqno 2@ 13
 \endcomment

We begin with the setup for geometric index theory~\cite{F}.  Let $\X\to S$
be a fiber bundle with fibers compact 4-manifolds.  Assume the vertical
tangent bundle $T(\X/S)\to \X$ is endowed with a spin structure and
Riemannian metric, and suppose too that there is a complementary horizontal
distribution.  Let $\WXS$~denote the curvature of the resulting Levi-Civita
covariant derivative on $T(\X/S)\to \X$.  We term $\X\to S$ with this data a
{\it Riemannian spin fiber bundle\/} or a {\it Riemannian spin manifold
over~$S$\/}.  Suppose $V\to\X$ is a hermitian vector bundle equipped with a
trivialization of~$\Det V\to\X$ and a compatible unitary covariant derivative
with curvature~$\WV$.  Said differently, $V$~and its covariant derivative are
associated to a principal $SU_N$~bundle with connection over~$\X$.  Recall
that spinor fields on a Riemannian spin 4-manifold are $\zt$-graded, the
grading termed `chirality', and the Dirac operator exchanges the chirality.
The geometric data determine a family of {\it chiral\/} Dirac operators~$\Dir
V$ parametrized by~$S$.  The chiral Dirac operators map positive chirality
$V$-valued spinor fields to negative chirality $V$-valued spinor fields.  Let
$\Det\Dir V\to S$ be the associated determinant line bundle with its natural
metric and covariant derivative.  It carries a natural $\zt$-grading as well:
the degree at~$s\in S$ is $\index D\mstrut _{\X_s}(V)\pmod2$.  We allow
$V$~to be a virtual bundle, or equivalently a $\zt$-graded bundle
$V=V^0\oplus V^1$, with both $\Det V^0$~and $\Det V^1$~trivialized.  In this
case we write
  $$ \Det \Dir V=\Det\Dir{V^0}\otimes \Det\Dir{V^1}^* \tag{2.1} $$
and $\rank V = \rank V^0 - \rank V^1$.

        \proclaim{\protag{2.2} {Theorem}}
 The  isomorphism class  of  the $\zt$-graded  determinant  line bundle  with
covariant derivative is
  $$ \bigl[\Det\Dir V\bigr] = \int_{\X/S}\cm(V)\qquad \text{in $\cE^2
     (S)$}. \tag{2.3} $$
        \endproclaim
\flushpar
 Recall that $\cm(V)\in \cE^6(\X)$ is the differential characteristic class
defined around~\thetag{1.16}; for a $\zt$-graded bundle $V=V^0\oplus V^1$ set
$\cm(V)=\cm(V^0) - \cm(V^1)$.  The integral
$\int_{\X/S}\:\cE^6(\X)\to\cE^2(S)$ uses the spin structure; see~\thetag{1.6}
and~\cite{HS,\S4.10}.  Also, note the splitting
  $$ \cE^2(S)\cong \cH^2(S)\times H^0(S;\zt) \tag{2.4} $$
under which a graded line bundle with covariant derivative maps separately to
the underlying line bundle with covariant derivative and the grading.  So
\theprotag{2.2} {Theorem} also determines the ungraded determinant line
bundle with covariant derivative.

A line bundle with covariant derivative is determined up to isomorphism by
all of its holonomies around loops.  In the language of differential
$E$-theory this is the following.

        \proclaim{\protag{2.5} {Lemma}}
 Let $L\to S$ be a $\zt$-graded complex line bundle with covariant derivative
and $[L]\in \cE^2(S)$ its isomorphism class.  Then $[L]$~is determined by the
grading of~$L$ in~$H^0(S;\zt)$ and by its integral over all loops~$\gamma
\:\cir\to S$, where $\cir$~has the bounding spin structure.  The integral
over such a loop is minus the log holonomy \rom(in~$\RR/\ZZ$\rom).
        \endproclaim

        \demo{Proof}
 By~\thetag{2.4} it suffices to show that the isomorphism class of the
underlying ungraded line bundle with covariant derivative is determined by
$\int_{S^1}\gamma ^*[L]$ for all~$\gamma $.  Since $S^1$~has the bounding
spin structure, we can write $\cir=\partial D^2$ as a spin manifold, and the
bundle with covariant derivative $\gamma ^*L\to\cir$ extends
to~$\widetilde{L}\to D^2$.  Let $\Omega ^{\widetilde{L}}$~denote its
curvature.  Then Stokes' theorem in differential $E$-theory implies that
  $$ \int_{S^1} \gamma ^*[L] \equiv \int_{D^2} \Omega ^{\widetilde{L}} \equiv
     -\log\hol_{S^1}(\gamma ^*L) \pmod1.  $$
That all such integrals determine the image of~$[L]$ in~$\cH^2(S)$ can be
seen directly from the exact sequence
  $$ 0 \longrightarrow H^1(S;\RR/\ZZ) \longrightarrow \cH^2(S)
     \longrightarrow \Omega _{\text{closed}}^2(S)  $$
in which the last arrow is the curvature: the curvature is determined by the
integral around loops $\gamma \:\cir\to S$ which bound a disk in~$S$, and the
integration map $H^1(S;\RR/\ZZ)\to \Hom\bigl(H_1(S),\RR/\ZZ \bigr)$ is an
isomorphism.
        \enddemo

The following lemma reduces the holonomy computation for~$\Det\Dir V$ to the
computation of its curvature. 

        \proclaim{\protag{2.6} {Lemma}}
 Suppose $Y$~is a closed spin 5-manifold and $V\to Y$ a rank~$N$ complex
vector bundle with~$c_1(V)=0$.  Then there exists a compact spin
6-manifold~$Z$ and a rank~$N$ complex vector bundle $W\to Z$ with $c_1(W)=0$
such that~$\partial Z=Y$ and~$W\res{\partial Z}\cong V$. 
        \endproclaim

\flushpar
 The lemma applies to virtual bundles as well: given $V=V^0\oplus V^1$ of
rank~$N$ and~$c_1(V)=c_1(V^0)-c_1(V^1)=0$ we can add trivial bundles to
replace~$V^1$ by a trivializable bundle and so $V^0$~by a bundle (of
rank~$\ge N$) with~$c_1(V^0)=0$.

        \demo{Proof}
 By Thom's theory this is the assertion that $\pi _5\,\MSpin\wedge (BSU_N)_+$
vanishes; here $X_+$~is the space~$X$ with disjoint basepoint.  First, we can
drop the~`$+$' since  
  $$ \pi _5\,\MSpin\wedge (BSU_N)_+ \cong \pi _5\,\MSpin\wedge BSU_N\;\times
     \;\pi _5\,\MSpin  $$
and $\pi _5\,\MSpin=0$.  Next, smash the cofiber sequence~\thetag{1.12}
with~$\MSpin$ to conclude from the long exact sequence in homotopy groups
that
  $$ \pi _5\,\MSpin\wedge BSU_N \cong \pi _5\,\MSpin\wedge (S^4\cup_\eta
     e_6)\;.  $$
Now smash the cofiber sequence 
  $$ S^4 \longrightarrow (S^4\cup_\eta e_6) \longrightarrow S^6
      $$
with~$\MSpin$ to obtain the exact sequence 
  $$ \pi _6\,\MSpin\wedge S^6 \longrightarrow \pi _5\,\MSpin\wedge S^4
     \longrightarrow \pi _5\,\MSpin\wedge (S^4\cup_\eta e_6)
     \longrightarrow \pi _5\,\MSpin\wedge S^6\;. \tag{2.7} $$
We have $\pi _6\,\MSpin\wedge S^6\cong \pi _0\,\MSpin\cong \ZZ$ and $\pi
_5\,\MSpin\wedge S^4\cong\pi_1\,\MSpin\cong \zt$.  Thus the initial map
in~\thetag{2.7} is identified with $\eta \:\pi _0\,\MSpin\to\pi _1\,\MSpin$,
which is surjective.  Since $\pi _5\,\MSpin\wedge S^6\cong \pi
_{-1}\,\MSpin=0$, it now follows that $\pi _5\,\MSpin\wedge (S^4\cup_\eta
e_6)=0$.
        \enddemo

        \demo{Proof of \theprotag{2.2} {Theorem}}
 First, the curvature of~$\Det\Dir V$ is the integral over~$\X\to S$ of the
6-form component of~$\Ahat(\WXS)\ch(\WV)$.  The flat trivialization of~$\Det
V$ implies $\Tr\WV=0$, and so the only contribution is
  $$ \int_{\X/S}ch_3(\WV) = \int_{\X/S}\frac 12\,c_3(\WV) =
     \int_{\X/S}\frac{-i}{48\pi ^3}\,\Tr(\WV)^3. \tag{2.8} $$
By~\thetag{1.14} this is the curvature of~$\int_{\X/S}\cm(V)$, i.e., the
differential form component of its image under $\cE^2(S)\to A^2(S)$
in~\thetag{1.8}.

 By \theprotag{2.5} {Lemma} to prove the theorem it suffices to verify that
the grading and the integral of both sides of~\thetag{2.3} over each loop
$\gamma \:\cir\to S$ agree.  The grading of~$\Det\Dir V$ at~$s\in S$ is given
by the index mod~2, which by the Atiyah-Singer index theorem and Rohlin's
theorem---the $\Ahat$~genus of a closed spin 4-manifold is even---is
  $$ \rank V\int_{\X_s}\Ahat(\X_s) -\int_{\X_s}-c_2(V) \equiv
     \int_{\X_s}\cm(V)\pmod2. \tag{2.9} $$
Now the pullback $Y=\gamma ^*\X\to\cir$ is a smooth 5-manifold, and it
obtains a spin structure from the spin structure of the fibers and the
bounding spin structure of the base~$\cir$.  \theprotag{2.5} {Lemma} implies
that the integral of~$[\Det\Dir V]$ is minus the log holonomy, which the
holonomy theorem for determinant line bundles~\cite{BF} computes as the
adiabatic limit of the Atiyah-Patodi-Singer invariant~$\xi _Y(\gamma ^*V)
\pmod1$.  The absence of~$\WXS$ in the curvature formula~\thetag{2.8}
implies that we can drop the adiabatic limit.  Let~$Y=\partial Z$ and $\gamma
^*V=\partial W$ as in \theprotag{2.6} {Lemma}.  Then the Atiyah-Patodi-Singer
index theorem implies
  $$ \xi _Y(\gamma ^*V)\equiv \int_{Z}\frac 12\,c_3(\Omega
     ^W)\pmod1. \tag{2.10} $$
On the other hand, by Stokes' theorem in differential $E$-theory
  $$ \int_{\cir}\int_{\gamma ^*\X/\cir}\cm(V) = \int_{Y}\cm(V) =
     \int_{Z}\wm(W) = \int_{Z}\frac 12\,c_3(\Omega ^W).  $$
The agreement with~\thetag{2.10} completes the proof. 
        \enddemo

If $V\to\X$ is either real or quaternionic, then the determinant bundle
simplifies. 

        \proclaim{\protag{2.11} {Corollary}}
 If $V\to\X$ has a quaternionic structure, compatible with its hermitian
structure and covariant derivative, then $\Det\Dir V$ has a real structure
compatible with its metric and covariant derivative.  Equation~\thetag{2.3}
holds and now $\cm(V)$ has order two by \theprotag{1.9(iii)} {Proposition}. 
        \endproclaim

\flushpar
 A quaternionic structure is a linear map $J\:V\to\bV$ with
$\overline{J}J=-\operatorname{id}_V$.  The spin space on a 4-manifold is also
quaternionic,\footnote{This implies the theorem of Rohlin used
in~\thetag{2.9}: the index of the Dirac operator is even since the kernel is
quaternionic.} whence the $V$-valued spinors are real.  The real structure
commutes with the Dirac operator, so induces a real structure on all
eigenspaces and so on the determinant bundle as well.  Note that the elements
of order two in~$\cE^2(S)$ form the abelian group $H^1(S;\zt)\times
H^0(S;\zt)$ of isomorphism classes of $\zt$-graded {\it real\/} line bundles
over~$S$.  
 
For the real case we have the following. 

        \proclaim{\protag{2.12} {Proposition}}
 If $V\to\X$ has a real structure compatible with its hermitian structure and
covariant derivative, then $\Det\Dir V$~is canonically trivial. 
        \endproclaim

\flushpar
 In this case the $V$-valued spinors are quaternionic, so the index is even
and hence the determinant line bundle has zero grading.  The patching
construction of the determinant line bundle~\cite{F} allows us to deduce its
triviality from the following lemma. 

        \proclaim{\protag{2.13} {Proposition}}
 Let $W$~be a hermitian vector space with compatible quaternionic structure.
Then $\Det W$~is canonically trivial. 
        \endproclaim

        \demo{Proof}
 The quaternionic structure $J\:W\to\overline{W}$ induces a real structure
$\det J\:\Det W\to\overline{\Det W}$ on the determinant line.  There are two
real points of norm one on~$\Det W$.  Let $e_1,\dots ,e_m$ be a unitary basis
of~$W$ over the quaternions.  Then 
  $$ e_1\wedge Je_1\wedge e_2\wedge Je_2\wedge \cdots\wedge e_m\wedge Je_m\in
     \Det W  $$
is a real point of norm one.  Since the space of such bases is connected,
this point in~$\Det W$ is independent of the basis. 
        \enddemo

 \head
 \S{3} Anomalies in Four-Dimensional Gauge Theories
 \endhead
 \comment
 lasteqno 3@ 10
 \endcomment

We begin in Minkowski spacetime~$M^4$.  Let $\SS$~be the two-dimensional
complex spin space, the half-spin representation of~$\Spin_{1,3}\cong
SL_2\CC$.  The opposite chirality spin space is its complex
conjugate~$\overline{\SS}$.  Let $H$~be a compact Lie group and $\rho
\:H\to\Aut(\VV^0)$ a unitary representation.  In a quantum field theory with
an $H$-gauge field~$A$, spinor fields which transform in the
representation~$\rho $ come in pairs\footnote{$\Pi \SS$~is the parity
reversal of~$\SS$, the $\zt$-graded vector space of degree one in which
spinor fields take values.}
  $$ \aligned
      \psi \:M^4&\longrightarrow \Pi \SS\otimes \VV^0 \\
      \bar\psi \:M^4&\longrightarrow \Pi \overline{\SS}\otimes
     \overline{\VV^0}\endaligned  $$
which are coupled in the Dirac lagrangian $\frac 12 \bar{\psi }\Dirac_A\psi
$.  The presence of spinor fields of both chiralities is dictated by
CPT-invariance.  It is natural to let $\VV=\VV^0\oplus \VV^1$ be a
$\zt$-graded representation with associated spinor fields
  $$ \aligned
      \psi \:M^4&\longrightarrow \Pi \SS\otimes (\VV^0\oplus
     \overline{\VV^1}) \\
      \bar\psi \:M^4&\longrightarrow \Pi \overline{\SS}\otimes
     (\overline{\VV^0}\oplus\VV^1)\endaligned  $$
Notice that $\VV$ and $\overline{\Pi \VV}=\overline{\VV^1}\oplus
\overline{\VV^0}$ lead to the same theory.
      
The Wick rotated Euclidean field theory is defined on the category of
Riemannian spin 4-manifolds~$X$.  The bosonic field is a connection~$\Theta $
on a principal $H$-bundle $P\to X$.  Gauge transformations---more generally,
isomorphisms of principal $H$-bundles---act as symmetries.  So the space
$\conn XH$ of $H$-connections on~$X$ must be considered as a groupoid, or as
a stack.\footnote{The local model is a smooth infinite dimensional manifold
with a smooth action of a compact Lie group.}  In any case we consider
families of connections parametrized by a smooth manifold~$S$ and allow
$X$~and its metric to vary as well.  In other words, we couple the gauge
theory to gravity, but we treat the gauge field and metric as classical
background fields.  Therefore, in the Euclidean theory we consider Riemannian
spin fiber bundles $\X\to S$, as in the beginning of~\S{2}, together with a
connection on a principal $H$-bundle $\sP\to\X$.  The representation $\rho
\:H\to\Aut(\VV)$ gives rise to an associated $\zt$-graded hermitian vector
bundle $V=V^0\oplus V^1\to \X$ with unitary covariant derivative.  Assume
that the fibers of $\X\to S$ are closed.  Then the Euclidean functional
integral over the spinor fields is
  $$ \det\Dir{V^0}\cdot \det\Dir{\overline{V^1}}\;:\;\X\longrightarrow
     \Det\Dir{V^0}\otimes \Det\Dir{\overline{V^1}}. \tag{3.1} $$
But since the spinors are self-conjugate, in fact quaternionic,
$\Dir{\overline{V^1}}=\overline{\Dir{V^1}}$ and so $\Det\Dir
{\overline{V^1}}\cong \overline{\Det\Dir{V^1}}\cong \Det\Dir{V^1}^*$.  Hence
the fermionic functional integral~\thetag{3.1} is a section of the
determinant bundle $\Det\Dir V\to\X$ defined in~\thetag{2.1} and computed in
\theprotag{2.2} {Theorem}.  
 
The next step in the quantum field theory is to perform an integral over the
space of bosonic fields, and one factor in the integrand is the fermionic
partition function~\thetag{3.1}.  To even set up the integral we need to
transform it from a section of a line bundle to a function, i.e., to take the
ratio with a trivialization $\triv\:S\to \Det\Dir V$.  We require that this
trivialization be geometric in the sense that $|\triv|=1$ and $\triv$~is
flat.  The {\it anomaly\/} is the obstruction to the existence of a flat
trivialization; it is measured by $[\Det\Dir V]\in \cE^2(S)$.  If the anomaly
vanishes, then there is a further requirement: the trivialization~$\triv$
must be consistent with gluing of 4-manifolds.  A consistent choice
of~$\triv$ is called a {\it setting of the quantum integrand\/}.  We remark
that the equivalence class of~$\Det\Dir V$ in~$\cE^2(S)$ includes its
$\zt$-grading.  We are unsure of the physical significance of this grading,
but believe that a theory is anomalous if the grading is nonzero.  Note in
this case that the fermionic functional integral~\thetag{3.1}~vanishes, since
the Dirac operator has a nonzero kernel.
 
\theprotag{2.2} {Theorem} applies to compute the anomaly if the determinant
of the representation $\rho \:H\to\Aut(\VV)$ is one.  If this condition does
not hold, then the determinant bundle is nontrivial in suitable families
where both the metric and gauge field vary: the curvature~\thetag{2.8} has
contributions from~$c_1(\WV)$.  Assume, then, that $\det\rho =1$.  Fix a
Riemannian spin 4-manifold~$X$.  Any $H$-connection on~$X$ is pulled back
from a universal connection on the classifying space~$BH$, so it suffices to
study the family of connections parametrized by $S=\Map(X,BH)$.  As at the
end of~\S{1} we construct a smooth model of~$BH$ and a universal connection.
The representation~$\rho $ determines a map $\hat\rho \: BH\to BSU_{N_1}\times
BSU_{N_2}$, where $\dim\VV^q=N_q$.  Let $\cm(\rho )\in \cE^6(BH)$ be the
pullback~$\hat\rho ^*(\cm_{N_1}-\cm_{N_2})$ of the universal differential
characteristic class.  Then if
  $$ e\:\Map(X,BH)\times X\longrightarrow BH  $$
is the evaluation map, \theprotag{2.2} {Theorem} computes the anomaly to be 
  $$ \int_{X}e^*\cm(\rho
     )=\int_{X}e^*\hat\rho^*(\cm_{N_1}-\cm_{N_2}). \tag{3.2} $$
The curvature of~$\cm(\rho )$, the 6-form on~$BH$ which is the pullback
of~\thetag{1.14} by~$\hat\rho $, is computed by applying the symmetric
trilinear form
  $$ \xi _1,\xi _2,\xi _3 \longmapsto \frac{-i}{96\pi ^3}\Tr \left[
     \drx1\drx2\drx3 + \drx2\drx1\drx3 \right],\qquad \xi _1,\xi _2,\xi _3\in
     \operatorname{Lie}(H), \tag{3.3} $$
on the Lie algebra of~$H$ to the curvature of the universal connection
on~$BH$.  This trilinear form is the usual expression for the {\it local\/}
anomaly in the physics literature, e.g.~\cite{We,\S22.3}.

If $\cm(\rho )$~vanishes, then there is no anomaly.  This in itself does not
provide a choice of trivialization of~$\Det\Dir V$, much less a consistent
choice under gluing---a setting of the quantum integrand.  For that we would
need a refinement of \theprotag{2.2} {Theorem} to an isomorphism of $\Det\Dir
V$ with an integral of a differential function representing~$\cm(V)$.  If
$\rho $~is a real representation, however, then \theprotag{2.12}
{Proposition} does provide a canonical trivialization.  

        \example{\protag{3.4 (The Standard Model)} {Example}}
 In this case $H=SU_3\times SU_2\times U_1$, or a finite quotient.  The
representation~$\rho $ is 15-dimensional, and it extends to the
representation~$\overline{V}\oplus {\wedge} ^2V$ of~$SU_5$, where
$V=\CC^5$~is the standard representation.  We compute $c_3{\tsize\wedge} ^2V=
c_3V$.  Hence $c_3(\overline{V}\oplus {\tsize\wedge} ^2V)=0$, and from
\theprotag{1.9(ii)} {Proposition} we have $\mu (\overline{V}\oplus
{\tsize\wedge} ^2V)=0$.  Thus $\cm (\overline{V}\oplus {\tsize\wedge} ^2V)=0$
as well.
      
An alternative argument: if we add a trivial representation to the
15-dimensional Standard Model representation~$\rho $, and then the sum
extends to the 16-dimensional half-spin representation of~$\Spin_{10}$.  We
claim $E^6(B\Spin_{10})=0$, whence $\cm(\rho )=0$ by~\thetag{1.11}.  The claim
follows from the long exact sequence~\thetag{1.3}; the fact that
$H^6(B\Spin_{10})=0$; and the fact that $\beta Sq^2w_4\in H^7(B\Spin_{10})$,
often denoted~$W_7$, is nonzero.
        \endexample

        \example{\protag{3.5 (The $SU_2$ anomaly~\cite{W3})} {Example}}
 Here $H=SU_2$ and $\rho $~is the standard representation, which is
quaternionic.  So by \theprotag{2.11} {Corollary} the determinant bundle is
real and the anomaly is of order~2; cf.~\theprotag{1.9(ii)} {Proposition}.
For~$X=S^4$, the case considered in~\cite{W3}, the anomaly is nonzero and is
more easily computed directly using Bott periodicity than from the
formula~\thetag{3.2} in terms of~$\cm$. 
        \endexample

The main theory of interest in this paper is QCD, the theory of quarks.  The
gauge group~$H=SU_{N_c}$, where $N_c$~is the number of ``colors''; in the
real world~$N_c=3$.  Let $\UU=\CC^{N_c}$ denote the fundamental
representation of~$H$.  There is another positive integer in the theory, the
number of ``flavors''~$N_f$.  In the real world there are six flavors of
quarks, but as only three of them are light in this context $N_f$~is often
taken to be equal to three.  Our discussion applies to any value of~$N_f$.
Let $\WW=\CC^{N_f}$.  The representation~$\rho $ of~$H$ is the $\zt$-graded
vector space
  $$ \VV=\UU\otimes \WW \,\;\oplus \,\; \UU\otimes \WW \tag{3.6} $$
where $H$~acts on $\UU$ as the fundamental representation and trivially
on~$\WW$.  The theory is trivially and canonically anomaly-free as
$\VV^0=\VV^1$.  QCD~has a {\it global\/} symmetry group $U_{N_f}\times
U_{N_f}$; the two factors independently act on the two copies of~$\WW$
in~\thetag{3.6}.  Our interest is the subgroup~$G\times G$, where
$G=SU_{N_f}$.  We digress now to briefly explain anomalies for global
symmetries and gauging of global symmetries.

Suppose we have a quantum field theory with global symmetry group a compact
Lie group~$K$.  If the space (stack) of fields on a manifold~$X$ is~$\field
X$, then $K$~acts on~$\field X$.  Let $\boson X$~be the stack of bosonic
fields, so there is a vector bundle $\field X\to\boson X$ with fibers the odd
vector spaces of fermions.  The group~$K$ acts on~$\boson X$ compatibly with
its action on~$\field X$.  The functional integral over the fermionic fields
is a $K$-invariant section of a $K$-equivariant $\zt$-graded hermitian line
bundle with covariant derivative on~$\boson X$.  The anomaly is the
obstruction to a {\it $K$-invariant\/} flat trivialization~$\triv$.  If
$K$~acts trivially on~$\boson X$, then it acts on the line bundle by a
character $K\to\TT$ on each component of~$\boson X$.  Here $\TT$~is the
circle group of unit norm complex numbers.  If the nonequivariant anomaly
vanishes, then the anomaly in the global symmetry is measured by these
characters.  This is the case in QCD, and for the subgroup $K=G\times
G=SU_{N_f}\times SU_{N_f}$ of the full global symmetry group~$U_{N_f}\times
U_{N_f}$ there are no nontrivial characters, whence no
anomalies.\footnote{The anti-diagonal $U_1\subset U_{N_f}\times U_{N_f}$ acts
by the character $\lambda \mapsto \lambda ^{m}$ for $m=-2N_cN_f\Sign(X)/8 -
2N_fk$ on the component where the second Chern class of the principal
$H=SU_{N_c}$-bundle is $k$~times the generator of~$H^4(X)$.  Here
$\Sign(X)$~is the signature of the spin manifold~$X$.  So this subgroup is
anomalous, though a finite cyclic subgroup is not.}
 
A {\it gauging\/} of the theory is an extension which includes a connection
on a principal $K$-bundle as a new field in the theory.  Thus in the gauged
theory the stack of fields $\tfield X$ on a manifold~$X$ fibers over the
stack~$\conn XK$ of $K$-connections on~$X$.  There is a distinguished point
of~$\conn XK$---the trivial connection with isotropy group~$K$---and we
require that the fiber of $\tfield X\to \conn XK$ at the trivial
$K$-connection be identified with the stack of fields~$\field X$ in the
original theory and the action of the isotropy group~$K$ on this fiber be the
original global symmetry.  Let $\tboson X$~denote the stack of bosonic fields
on~$X$ in the gauged theory.  Then the fermionic anomaly in the gauged
theory, which is the isomorphism class of a $\zt$-graded line bundle with
covariant derivative over~$\tboson X$, restricts on the fiber~$\boson X$ over
the trivial $K$-connection to the anomaly in the original theory (including
the global $K$-action).
 
QCD~has a natural extension which gauges the global $G\times G=SU_{N_f}\times
SU_{N_f}$~symmetry.  It is a four-dimensional gauge theory with gauge group
$H\times G\times G=SU_{N_c}\times SU_{N_f}\times SU_{N_f}$.  The
representation~$\rho $ acts on the vector space~\thetag{3.6}: the
group~$SU_{N_c}$ acts on both copies of~$\UU$ as before, the first factor
of~$SU_{N_f}$ acts on the first copy of~$\WW$, and the second factor
of~$SU_{N_f}$ acts on the second copy of~$\WW$.  The stack of bosonic fields
in the gauged theory on a fixed 4-manifold~$X$ is $\tboson X=\conn XH \times
\conn X{G\times G}$.  As usual, we consider smooth families of fields, which
for the gauged theory is a Riemannian spin fiber bundle $\X\to S$ with
compact 4-manifolds as fiber and a principal $SU_{N_c}\times (G\times
G)$~bundle $\sP\times \sQ\to\X$ with connection.  Let
$\cm_1(\sQ),\cm_2(\sQ)\in \cE^6(\X)$ be the differential characteristic
classes associated with the two factors of~$G$.

        \proclaim{\protag{3.7} {Proposition}}
 The anomaly in gauged QCD is  
  $$ \int_{\X/S}N_c\bigl(\cm_1(\sQ) - \cm_2(\sQ)\bigr). \tag{3.8} $$
        \endproclaim

\flushpar
 This follows directly from \theprotag{2.2} {Theorem}, where we use
\theprotag{1.9(v)} {Proposition} to compute the characteristic class of the
vector bundle associated to the representation~\thetag{3.6}.  (Recall that
$G\times G$~acts trivially on~$\UU$, which has dimension~$N_c$.)  In terms of
the stack~$\tboson X$ of bosonic fields, the anomaly is pulled back
from~$\conn X{G\times G}$ and is given by~\thetag{3.8}.

The symmetry breaking in QCD is deduced from expectation values of bilinear
expressions in the spinor fields.  If $(\psi _1,\psi _2)$ are the spinor
fields corresponding to the representation~\thetag{3.6}, then they have the
form $\langle \overline{\psi _1},T\cdot \psi _2 \rangle$, where the inner
product is that in $\UU\otimes \WW$ and $T$~is an element of the Lie algebra
of~$SU_{N_f}$ which acts on $\UU\otimes \WW$ as the identity on~$\UU$ tensor
its action on~$\WW$.  The expectation value is taken at any point of
Minkowski spacetime, as it is constant by Poincar\'e invariance.  To write
this bilinear in the gauged theory, and so implement the symmetry breaking,
we need additional data.  Namely, there are vector bundles~$W_1,W_2$
associated to the two $G$-connections, and we need an isomorphism~$W_1\cong
W_2$, so an isomorphism of the principal $G$-bundles underlying the two
$G$-connections.  The construction of the gauged effective theory in~\S{4}
includes that isomorphism; in fact, it is the scalar field in that theory.

Gauged~QCD admits new topological terms.  In the {\it exponentiated\/} action
these have the form 
  $$ \exp\left( 2\pi i\theta _1\int_{\X/S}c_2(\sQ)_1 + 2\pi i\theta
     _2\int_{\X/S}c_2(\sQ)_2 \right), \tag{3.9} $$
where $c_2(\sQ)_1,c_2(\sQ)_2$~are the degree four characteristic classes
corresponding to the two $G=SU_{N_f}$~factors, and $\theta _1,\theta _2\in
\RR/\ZZ$.  The requirement that a fermion bilinear exist implies that
$c_2(\sQ)_1=c_2(\sQ)_2$, as argued in the previous paragraph, so we have a
single topological term with coefficient~$\theta =\theta _1+\theta _2$. 

Although the gauging of~$\GG$ leads to an anomaly, there are subgroups which
can be gauged to give a viable theory.

        \definition{\protag{3.10} {Definition}}
 A subgroup $\iota \:K\hookrightarrow G\times G$ is called {\it
anomaly-free\/} if $(B\iota )^*(\mu _1-\mu _2)\in E^6(BK)$ vanishes.
        \enddefinition

\flushpar
 If $K$~is anomaly-free, then it follows from \theprotag{3.7} {Proposition}
that the theory obtained by gauging the global symmetry group~$K$ is
anomaly-free.  Any subgroup of~$G$ embedded diagonally in~$G\times G$ is
clearly anomaly-free.  For the subgroup $\iota \:SU_2\times
\{1\}\hookrightarrow G\times G$ the pullback $(B\iota )^*(\mu _1-\mu _2)$ is
torsion of order two.  Thus, even though not detected rationally, this
subgroup is not anomaly-free.  (Compare~\cite{W1,p.~431}.)

 \head
 \S{4} The Wess-Zumino-Witten Term in the Low Energy Theory of Pions
 \endhead
 \comment
 lasteqno 4@ 18
 \endcomment

The low energy dynamics of the pions is described by a $\sigma $-model with
target~$(\GG)/G$, as explained in the introduction.  Here, as before,
$G=SU(N)$ with~$N\ge2$.  The kinetic and mass terms have the usual form.
Here we give a novel definition of the topological term in the action.  As
this Wess-Zumino-Witten term is only determined up to integer shifts, we work
with its exponential in the exponentiated Euclidean
action~$e^{-S_{\text{Eucl}}}$.  The space of fields in the $\sigma $-model on
a manifold~$X$ is $\field X=\Map(X,G)$.

        \definition{\protag{4.1} {Definition}}
 Let $X$~be a closed spin 4-manifold.  The {\it WZW factor\/} evaluated on
$\phi \:X\to G$ is
  $$ W_X(\phi ) = \exp\left( 2\pi i\int_{X}N_c\,\phi ^*\cn
     \right). \tag{4.2} $$
        \enddefinition

\flushpar
 Recall that $\cn\in \cE^5(G)$ is the differential $E$-class defined in
\theprotag{1.9(iv)} {Proposition} and~\thetag{1.15}.  Note that the integral
$\int_{X}\:\cE^5(X)\to \cE^1(\point)$ takes values in~$\RR/\ZZ$, as follows
immediately from~\thetag{1.7}.  So $W_X(\phi )$~is a well-defined element
of~$\CC$ with unit norm.  The factor of~$N_c$ is put to match the high energy
theory; see \theprotag{4.17} {Proposition} below.  Since $\cn$~is invariant
under left and right translations by~$G$, the WZW-factor~\thetag{4.2} is also
$(G\times G)$-invariant.  If $X=\partial Z$~is the boundary of a compact {\it
spin\/} 5-manifold~$Z$, and $\phi \:X\to G$ extends to $\Phi \:Z\to G$, then
Stokes' theorem for differential $E$-theory implies
  $$ W_X(\phi )= \exp\left( 2\pi iN_c\int_{Z}\Phi ^*\wn \right) = \exp\left(
     N_c\int_{Z}\frac{1}{240\pi ^2}\Tr(\Phi ^*\theta )^5 \right), \tag{4.3} $$
which is the usual formula in the physics literature.  (Recall that $\theta
=g\inv dg$ is the Maurer-Cartan form on~$G=SU_{N_f}$.) The signature of~$X$
is an obstruction to the existence of~$Z$, so \thetag{4.3} cannot serve as a
definition of the WZW~factor.\footnote{However, the square~$W_X(\phi )^2$ is
expressed in terms of ordinary differential cohomology, and since $H_4(G)=0$
it may defined via an integral of a differential form over a bounding
5-chain.} 
 
We proceed to verify properties~(i)--(iv) from the introduction.  To
demonstrate that \thetag{4.2}~depends on a spin structure, we compute the
dependence explicitly.   

        \proclaim{\protag{4.4} {Proposition}}
 Let $\delta \in H^1(X;\zt)$ be the difference between two isomorphism
classes of spin structures on~$X$ with the same underlying orientation.
Then the ratio of the WZW factors~\thetag{4.2} computed with the two spin
structures is~$\pm1$ according to the value of
  $$ N_c\,\delta \smile \phi ^* \overline{h_3}[X] \quad \in \quad \zt,
     \tag{4.5} $$
where $\overline{h_3}$ is the nonzero element of~$H^3(G;\zt)$ and $[X]\in
H_4(X)$ is the fundamental class. 
        \endproclaim

        \demo{Proof}
 Let $e$~denote the Anderson dual theory~\cite{HS,Appendix~B},
\cite{FMS1,Appendix~B} to~$E$.  Then $e^0(\point)\cong \ZZ$,
$e^{-1}(\point)\cong \zt$, and all other groups vanish.  The spectrum fits
into a fibration 
  $$ \Sigma H\zt \longrightarrow e\longrightarrow H\ZZ \tag{4.6} $$
whose $k$-invariant is the nontrivial map 
  $$ Sq^2\circ r\:H\ZZ\to\Sigma ^2H\zt, \tag{4.7} $$
the composition of the Steenrod square with reduction mod~2.  The cohomology
theory~$e$ is {\it multiplicative\/}, that is, $e$~is a ring spectrum.  This
can be seen in several ways.  First, the cohomology class represented
by~\thetag{4.7} is {\it primitive\/}, and so it follows that the fiber~$e$ is
a ring spectrum.  We can also identify~$e$ as a Postnikov truncation of
connective $ko$-theory, and again it follows that $e$~is a ring.  More
concretely, the zero space is the classifying space of the category of
$\ZZ$-graded real lines, and the latter is a ring: addition is the tensor
product of lines, and the multiplication of $L_1,L_2$ in degrees~$d_1,d_2$
puts $L_1^{\otimes d_2}\otimes L_2^{\otimes d_1}$ in degree~$d_1d_2$.  One
can also see~$e$ as a truncation of the sphere spectrum, so identify its
points as framed 0-manifolds; the sum is then disjoint union and the product
is Cartesian product.  It follows that~$E$, the Anderson dual of~$e$, is a
module over the ring~$e$.  
 
Suppose $V\to X$ is a real vector bundle of rank~$N$ over a space~$X$.  Then
\thetag{4.6}~leads to the long exact sequence of cohomology groups 
  $$ \xymatrix{ \cdots \ar[r] &H^{N+1}(V;\zt)\cv \ar[r]^(.6){s} &e^N(V)\cv
     \ar[r] &H^N(V;\ZZ)\cv \ar[r]^(.4){Sq^2\circ r} &H^{N+2}(V;\zt)\cv \ar[r]
     &\cdots \\
      & H^1(X;\zt)\ar[u]^{\cong }_{\bU} &&& H^2(X;\zt)\ar[u]^{\cong }_{\bU}}
     \tag{4.8} $$
Here `cv' denotes compact vertical supports and the vertical arrows are Thom
isomorphisms.  Assume $V$~is oriented with Thom class $U\in H^N(V;\ZZ)\cv$,
and let $\bU\in H^N(V;\zt)\cv$ be the mod~2 Thom class.  Then $(Sq^2\circ
r)(U)=\bU w_2(V)$, where $w_2(V)\in H^2(X;\zt)$ is the second Stiefel-Whitney
class.  Hence a spin structure on~$V$---a trivialization
of~$w_2(V)$---induces a lift~$U_e\in e^N(V)\cv$ of~$U$, a Thom class in
$e$-theory.  Lifts differ by~$\bU\delta $ for~$\delta \in H^1(X;\zt)$. 
 
Turning to the proposition, let $\pi \:V\to X$ be the normal bundle to an
embedding $X\hookrightarrow \RR^{N+4}$, let $U_e$~be the $e$-Thom class for
some spin structure, and $\cUe$ a lift to a differential Thom class in
$e$-theory~\cite{HS}.  The integral of a class~$\cb\in \cE^5(X)$ is computed
as the product
  $$ \cUe\cdot \pi ^*\cb\quad \in\quad \cE^{N+5}(\RR^{N+4})\cong \RR/\ZZ. $$
If $\delta \in H^1(X;\zt)$ is a change of spin structure, then $U_e$~changes
by~$s(\bU\delta )$ in the sequence~\thetag{4.8}.  Now there is an inclusion 
  $$ \xymatrix{e^{N-1}(V;\RZ)\cv \;\ar@{^{(}->}[r]^(.6)f &\;\ce^N(V)\cv} $$
of ``flat'' elements in differential $e$-theory.  Smash~\thetag{4.6} with
the Moore space for~$\RZ$ to construct a short exact
sequence\footnote{The degree shift in the first term is explained with chain
complexes: $\ZZ\to \RR$ with $\ZZ$~in degree~$-1$ is quasi-isomorphic
to~$\RZ$ in degree zero}  
  $$ 0 @>>> H^{N+1}(V;\zt)\cv @>s_{\RZ}>> e^{N-1}(V;\RZ)\cv @>>>
     H^{N-1}(V;\ZZ)\cv @>>> 0. $$
The change in spin structure shifts the differential Thom class~$\cUe$ by the
image of~$\bU\delta $ under the composition~$f\circ s_{\RZ}$.  Then the
change in the product~$\cUe\cdot \pi ^*\cb$ depends only on the
image~$j(\beta )$ of~$\cb$ under $\cE^5(X) @>>> E^5(X)@>j>> H^3(X;\zt)$: it
is $f\bigl(\bU\delta \cdot \pi ^*j(\beta ) \bigr)\in
\cE^{N+5}(\RR^{N+4})\cong \RZ$.  We claim that this equals
  $$ \delta \smile j(\beta )\quad \in \quad H^4(X;\zt) \cong \zt \cong \tfrac
     12\ZZ/\ZZ \subset \RZ, $$
which leads immediately to~\thetag{4.5}.  
 
The claim amounts to showing that the composition $ \Sigma ^2H\zt \wedge
E\longrightarrow e_{\RZ}\wedge E \longrightarrow E_{\RZ} $ is nonzero.  By
shifting the Moore space for~$\RZ$ in the wedge, this is equivalent to
showing that the module map $e\wedge E_{\RZ}\to E_{\RZ}$ induces the nonzero
map $\pi _1e \otimes \pi _{-1}E_{\RZ}\to \pi _0E_{\RZ}$.  But since
$E_{\RZ}$~ is the Pontrjagin dual of~$e$, this is obtained by
applying~$\Hom(-,\RZ)$ to the multiplication $\pi _1e\: \pi _0e\to \pi _1e$,
and the latter is nonzero.
        \enddemo

We now turn to property~(ii) in the introduction.  Let $Y$~be a compact spin
3-manifold, and consider canonical quantization of the $\sigma $-model
on~$Y$.  The space of classical solutions of the $\sigma $-model
on~$\RR\times Y$ with its Lorentz metric is the space of solutions to a wave
equation, which (at least formally) is the space of Cauchy data.  The latter
is identified with the tangent bundle of~$\Map(Y,G)$.  It carries a
symplectic structure, and the zero-section $\Map(Y,G)$ is lagrangian.
Without the WZW~factor the Hilbert space would, at least formally, be the
space of $L^2$~functions on~$\Map(Y,G)$.  In particular, it would only
consist of bosonic states.  The WZW~factor changes the symplectic structure
of~$T\!\Map(Y,G)$ in a geometric manner: the normalized curvature of a
hermitian line bundle with covariant derivative pulled back from~$\Map(Y,G)$
is added.  Let
  $$ e\:\Map(Y,G)\times Y\longrightarrow G  $$
be the evaluation map.  Then the isomorphism class of this line bundle may be
written 
  $$ \int_{Y}N_c\, e^*\cn\quad \in \quad \cE^2\bigl(\Map(Y,G)
     \bigr). \tag{4.9} $$
The important point for us is that $\cE^2$~parametrizes {\it $\zt$-graded\/}
line bundles with covariant derivative.  The quantum Hilbert space is now the
space of sections of this line bundle, so it too is $\zt$-graded.  The
$\zt$-grading of the quantum Hilbert space reflects statistics of states:
even degree states are bosonic and odd degree states are fermionic.  More
precisely, the grading of~\thetag{4.9} at $\phi \in \Map(Y,G)$ is
  $$ \int_{Y}N_c\,\phi ^*j(\nu ) \equiv \int_{Y}N_c\,\phi ^*h_3 \equiv
     N_c\, \deg_2(\phi )\quad \in \quad \zt, \tag{4.10} $$
where $\deg_2(\phi )$~is the mod~2 degree of $\phi \:Y\to G$, the homology
class of~$\phi $ in~$H_3(G;\zt)\cong \zt$.  (Recall that $h_3\pmod2$ is the
generator of~$H^3(G;\zt)$; cf.~\thetag{1.10}.)  In physical term the integer
degree is identified with the {\it baryon number\/}, and \thetag{4.10}---the
statistics formula for solitons which is an immediate consequence of
\theprotag{4.1} {Definition}---matches the formula derived from
physics~\cite{W2}.
 
Property~(iii) of the introduction is evident: our definition works
for~$G=SU(2)$ since $\cn$~is a nonzero element of order two and
\thetag{4.10}~is still valid.  So the WZW~factor encodes the statistics of
solitons.  
 
The remainder of this section is devoted to property~(iv).  To that end we
construct an extension of the $\sigma $-model with WZW~factor~\thetag{4.2}
when the global $(G\times G)$~symmetry is gauged.  (See the discussion
preceding \theprotag{3.7} {Proposition} for generalities on gauging.)  As the
gauged extensions of the kinetic and mass terms are straightforward and
anomaly-free, we only consider the WZW~factor.  The fields~$\tfield X$ in the
gauged theory on a Riemannian spin 4-manifold~$X$ are a connection~$\Theta $
on a principal $(G\times G)$-bundle $Q\to X$ and a $(G\times G)$-equivariant
map $\phi \:Q\to G$, i.e., a section~$\phi $ of the associated fiber bundle
$G_Q=Q\times _{(G\times G)}G\to X$ with fiber~$G$.  Note the fibering
$\tfield X\to \conn X{G\times G}$ as required by the general theory.  If the
class $\cn\in \cE^5(G)$ extended to a class in~$\cE^5(G_Q)$, then
\thetag{4.2}~would be valid with this extended class replacing~$\cn$, and
there would be an anomaly-free gauging of the WZW~factor.  But such an
extension does not exist, and so the gauged theory is more subtle---and
anomalous.
 
Even the topological class $\nu \in E^5(G)$ does not extend, and to measure
the obstruction we work universally.  Let $\sE\to B(G\times G)=BG\times BG$
be the fiber bundle 
  $$ \pi \:\sE=G_{E(G\times G)}=E(G\times G)_{(G\times G)}G\longrightarrow
     B(G\times G). \tag{4.11} $$
The bundle~$\sE$ is the quotient of $EG\times EG$ by the diagonal $G$-action,
so is homotopy equivalent to~$BG$; the projection to~$BG\times BG$ is
homotopy equivalent to the diagonal $BG\to BG\times BG$.  Recall that
ordinary cohomology is defined in terms of {\it cochains\/} and a {\it
differential\/}.  A {\it cocycle\/} is a cochain whose differential vanishes,
and the cohomology is the quotient of cocycles by differentials of cochains.
The entire theory is $\ZZ$-graded and the differential has degree one.  In
generalized cohomology there are analogous notions, and for now we simply
call them `E-cochains', etc.; in~\S{5} we give proper definitions.

        \proclaim{\protag{4.12} {Theorem}}
 There is an $E$-cochain~$\alpha $ of degree~5 on the total space~$\sE$
of~\thetag{4.11} which satisfies:\newline 
 \indent{\rm (i)}\ the restriction of~$\alpha $ to a fiber has zero
$E$-differential and represents~$\nu \in E^5(G)$;\newline 
 \indent{\rm (ii)}\ the $E$-differential of~$\alpha $ is the pullback of an
$E$-cocycle of degree~6 on~$B(G\times G)$ which represents $\mu _1-\mu _2\in
E^6\bigl(B(G\times G) \bigr)$.   
        \endproclaim

\flushpar
 We term~$\alpha $ a {\it transgressing $E$-cochain\/} and say that $\nu
$~transgresses to~$\mu _1-\mu _2$.  Transgression in ordinary cohomology is
related to the Leray-Serre spectral sequence~\cite{BT,\S18}, but that tool is
not available for $E$-cohomology.  \theprotag{4.12} {Theorem} has a
restatement in terms of Borel equivariant $E$-cohomology: the $E$-cohomology
class~$\nu \in E^5(G)$ does {\it not\/} has an equivariant extension
to~$E^5_{G\times G}(G)$.  The existence of such an extension is obstructed by
$\mu _1-\mu _2\in E^6_{G\times G}(\point)$.  Then $\alpha $~may be regarded
as a $(G\times G)$-equivariant $E$-cochain on~$G$.  We defer the proof of
\theprotag{4.12} {Theorem} and further discussion to~\S{5}.
 
To define the gauged WZW~factor we need an extension of \theprotag{4.12}
{Theorem} to differential $E$-theory. 

        \proclaim{\protag{4.13} {Theorem}}
 There is an $\cE$-cochain~$\ca $ of degree~5 on the total space~$\sE$
of~\thetag{4.11} which satisfies:\newline 
 \indent{\rm (i)}\ the restriction of~$\ca $ to a fiber has zero
$\cE$-differential and represents~$\cn \in \cE^5(G)$;\newline 
 \indent{\rm (ii)}\ the $\cE$-differential of~$\ca $ is the pullback of an
$\cE$-cocycle of degree~6 on~$B(G\times G)$ which represents $\cm _1-\cm _2\in
\cE^6\bigl(B(G\times G) \bigr)$.    
        \endproclaim

\flushpar
 The proof is in~\S{5}.  The main idea in \theprotag{4.13} {Theorem} beyond
\theprotag{4.12} {Theorem} is that the differential form~$\wm$ is a
transgression of~$\wn$ in the universal bundle $EG\to BG$.  The transgressing
form is a 5-form on~$EG$, the Chern-Simons form.  We then construct a
5-form~\thetag{5.16} on~$\sE$ whose de Rham differential is the pullback of
$\omega _{\cm_1}-\omega _{\cm_2}\in \Omega ^6\bigl(B(G\times G) \bigr)$.  In
fact, there are different equivalence classes of transgressing $\cE$-cochains
$\ca$ which differ by the inclusion of a topological term in the gauged
WZW~model.
 
We need one more maneuver to define the gauged WZW~factor.  Recall that
$\conn X{\GG}$ is a stack represented by the groupoid~$\sG_1$ in which an
object~$(Q,\Theta )$ is a connection~$\Theta $ on a principal $(\GG)$-bundle
$Q\to X$.  A morphism $(Q,\Theta )\to (Q',\Theta ')$ is a $\GG$-equivariant
map $\varphi \:Q\to Q'$ such that $\varphi ^*\Theta '=\Theta $.  We
replace~$\sG_1$ by an equivalent groupoid~$\sG_2$; it also represents the
stack~$\conn X{\GG}$.  An object~$(Q,\Theta ,\gamma )$ in~$\sG_2$ is a triple
where $\gamma \:Q\to E(\GG)$ is a $(\GG)$-equivariant map which classifies
the connection~$\Theta $ on~$Q$.  In other words, $\gamma ^*\Tu=\Theta $,
where $\Tu$~is the universal connection on $E(\GG)\to B(\GG)$.  (As in the
construction at the end of~\S{1}, we can avoid the condition~$\gamma
^*\Tu=\Theta $ by including a Chern-Simons term.)  Morphisms in~$\sG_2$ are
described at the end of~\S{5}.  The important point is that the space of
classifying maps~$\gamma $ for fixed~$(Q,\Theta )$ is contractible and
nonempty.  In essence we adjoin this contractible choice as a new field and
posit a symmetry which makes it inessential (auxiliary).
 
A field in the gauged $\sigma $-model is, therefore, a principal
$(\GG)$-bundle $Q\to X$ with connection~$\Theta $, a classifying map $\gamma
\:Q\to E(\GG)$ for~$\Theta $, and a section $\phi$ of the associated bundle
$G_Q\to X$ with fiber~$G$.  The classifying map~$\gamma $ induces a
classifying map $\tg\:G_Q\to\sE$.   

        \definition{\protag{4.14} {Definition}}
 The gauged WZW factor is 
  $$ \tW_X(Q,\Theta ,\gamma ,\phi ) = \exp\left(\tpi \int_{X}N_c\ \phi
     ^*\tg^*\ca \right) . \tag{4.15} $$
        \enddefinition

\flushpar
 The $\cE$-cochain $\tg^*\ca$ on~$G_Q$ is the gauged extension of the
cocycle~$\cn$ on~$G$, so \thetag{4.15}~is a natural generalization
of~\thetag{4.2}.  We discuss the well-definedness of this definition at the
end of~\S{5}.
 
To analyze this definition we work in smooth families.  Let $\X\to S$ be a
Riemannian spin fiber bundle with compact 4-manifolds as fibers, $\sQ\to\X$ a
principal $(\GG)$-bundle with connection~$\Theta $, $\gamma \:\sQ\to E(\GG)$ a
classifying map for the connection, and $\phi $~a section of $G_{\sQ}\to\X$.
The gauged WZW~factor is now an integral over the fibers
  $$ \tW_X(\sQ,\Theta ,\gamma ,\phi ) = \exp\left(\tpi \int_{\X/S}N_c\ \phi
     ^*\tg^*\ca \right), \tag{4.16} $$
and the result is an $\cE$-cochain on~$S$ of degree~1.  Its
$\cE$-differential is an $\cE$-cocycle of degree~2, so represents a class
in~$\cE^2(S)$.  Now from \theprotag{4.13(ii)} {Theorem} the $\cE$-differential
of the integrand represents the $\cE$-cohomology
class~$N_c\,\bg^*(\cm_1-\cm_2)$, and so by Stokes' theorem the
$\cE$-differential of~\thetag{4.16} is
  $$ \int_{\X/S}N_c\,\bg^*(\cm_1-\cm_2) =
     \int_{\X/S}N_c\bigl(\cm_1(\sQ)-\cm_2(\sQ) \bigr),  $$
in terms of the differential characteristic class~$\cm$ defined
around~\thetag{1.16}.  Recall that an $\cE$-cocycle of degree~2 may be
represented as a $\zt$-graded hermitian line bundle with unitary covariant
derivative.  Then a $\cE$-cochain of degree~1 whose differential is that
cocycle may be represented as a not-necessarily-flat section of this bundle
of unit norm.  That section~\thetag{4.15} is part of the gauged $\sigma
$-model action, and therefore the line bundle is the anomaly.  This proves
the following.

        \proclaim{\protag{4.17} {Proposition}}
 The anomaly in the gauged $\sigma $-model with WZW~factor is 
  $$ \int_{\X/S}N_c\bigl(\cm_1(\sQ)-\cm_2(\sQ) \bigr). \tag{4.18} $$
        \endproclaim

\flushpar
 The agreement of~\thetag{3.8} and~\thetag{4.18} is the 't Hooft anomaly
matching.  In terms of the stack of (bosonic) fields, which is a fibering
$\tfield X\to\conn X{\GG}$ over the stack of $(\GG)$-connections, the
classical action with gauged WZW~factor is a section of a line bundle
over~$\tfield X$, and that line bundle is pulled back from~$\conn X{\GG}$.  It
is the anomaly in the gauged $\sigma $-model, and its isomorphism class is
computed by \theprotag{4.17} {Proposition}.  This completes the verification
of property~(iv) of the introduction.

The existence of a section~$\phi $ of $G_{\sQ}\to \X$ implies a topological
restriction on the $\GG$~bundle $\sQ\to\X$, namely that the two constituent
$G$-bundles $\sQ_1\to\X$ and $\sQ_2\to\X$ be isomorphic as topological
principal bundles.  (A section of the associated $(\GG)/G$~bundle is
equivalent to a reduction of structure group of~$\sQ$ to the diagonal
$G\subset G\times G$.)  This is precisely the condition in QCD to define the
fermion bilinear; see the discussion preceding~\thetag{3.9}.  It also implies
that the {\it topological\/} anomaly in \theprotag{4.17} {Proposition}
vanishes, but there may still be a {\it geometric\/} anomaly: the
$G$~bundles~$\sQ_1,\sQ_2$ need not be isomorphic as bundles with connection.

  \head
 \S{5} Transgression
 \endhead
 \comment
 lasteqno 5@ 18
 \endcomment

We recommend~\cite{DK,\S6} as an introduction to the topology used in this
section.

The cochains used to define ordinary cohomology are replaced in a generalized
cohomology theory~$E$ by maps into the representing spectrum~$\{E_n\}$.
Recall that $E_n$~is a pointed topological space and the spectrum comes
equipped with maps $\Sigma E_n\to E_{n+1}$.  An `$E$-cocycle' (as used
in~\S{4}) of degree~$n$ on a topological space~$X$ is simply a map $X\to
E_n$.  Homotopic maps are considered equivalent, and the $E$-cohomology
group~$E^n(X)$ is the set of homotopy classes~\thetag{1.1}.  An `$E$-cochain'
of degree~$n$ on~$X$ is a {\it based\/} map 
  $$ CX\longrightarrow E_{n+1} \tag{5.1} $$
from the {\it unreduced\/} cone\footnote{The unreduced cone on~$X$ is the
pointed space
  $$ CX = [0,1]\times X\bigm/ \{1\}\times X  $$
with basepoint~$\{1\}\times X$.  Note $X\subset CX$ as $\{0\}\times X$ and
$CX/X$~is the {\it unreduced\/} suspension~$\Sigma X$.  As $X$~does not have
a basepoint, the notation for suspension is unambiguous.} on~$X$; its
`$E$-differential' is the restriction to~$X\subset CX$.  If that restriction
is trivial---maps to the basepoint~$*$ of~$E_{n+1}$---then
\thetag{5.1}~factors to a based map $\Sigma X\to E_{n+1}$, and by adjunction
it is equivalent to a map $X\to \Omega E_{n+1}\simeq E_n$, so represents a
class in~$E^n(X)$.  If $A\subset X$ is a subspace, then a class in the
relative cohomology group~$E^n(X,A)$ is represented by a map $X\cup CA\to
E_n$.

We now define transgression in generalized cohomology.

        \definition{\protag{5.2} {Definition}}
 Let $F @>i>> \sE @>\pi >>B$ be a fibration.  Then $\nu \in E^n(F)$ is
{\it related by transgression\/} to $\mu \in E^{n+1}(B)$ if there exists
$\sigma \in E^{n+1}(\sE,F)$ and $\mu _0\in E^{n+1}(B,b_0)$ such that under
the maps 
  $$ \xymatrix@1{E^n(F)\ar[r]^-\delta &E^{n+1}(\sE,F) &
     E^{n+1}(B,b_0)\ar[l]_<<<{\pi ^*} \ar[r]^-j & E^{n+1}(B)} \tag{5.3} $$
we have $\sigma =\delta (\nu) =\pi ^*(\mu _0)$ and $\mu =j(\mu _0)$. 
        \enddefinition

\flushpar
 All spaces are pointed; the basepoint of~$B$ is~$b_0$, and $\pi \inv
(b_0)=F$ is the fiber of~$\pi $.  The map~$\delta $ is the connecting
homomorphism in the long exact sequence of the pair~$(\sE,F)$, and $j$~is the
homomorphism which forgets the basepoint.\footnote{$E^{\bullet}(B,b_0)$~is
the {\it reduced\/} $E$-cohomology of~$B$, often denoted
$\widetilde{E}^\bullet(B)$.}  The set of transgressive elements in~$E^n(F)$
forms a subgroup, and transgression is only well-defined into a quotient
of~$E^{n+1}(B)$.  The relation of \theprotag{5.2} {Definition} with the
description in \theprotag{4.12} {Theorem} is the following.

        \proclaim{\protag{5.4} {Lemma}}
 Let $F @>i>> \sE @>\pi >>B$ be a fibration.  Then $\nu \in E^n(F)$ is
related by transgression to~$\mu \in E^{n+1}(B)$ if and only if there exists
a map $\alpha \:C\sE\to E_{n+1}$ such that\newline
 \indent{\rm (i)}\ the restriction of~$\alpha $ to $F\subset CF\subset C\sE$
is trivial and the restriction of~$\alpha $ to~$CF$ factors through a map
$\Sigma F\to E_{n+1}$ which represents~$\nu $;\newline
 \indent{\rm (ii)}\ the restriction of~$\alpha $ to~$\sE\subset C\sE$ is the
pullback under~$\pi $ of a {\it based\/} map $B\to E_{n+1}$ which
represents~$\mu $. 
        \endproclaim

        \demo{Proof}
 The class~$\sigma $ in \theprotag{5.2} {Definition} is represented by a map
$c\:\sE\cup CF\to E_{n+1}$.  Because~$\sigma =\delta (\nu )$, it extends to
$\tilde{c}\:C\sE\cup CF\simeq\Sigma F\to E_{n+1}$, and the extension
represents~$\nu $.  Also, because~$\sigma =\pi ^*(\mu _0)$, the restriction
of~$c$ to~$\sE$ is pulled back from a based map $G\to E_{n+1}$, and in
particular is trivial on~$F\subset \sE$.  By a homotopy we can assume that
$c$~is trivial on~$CF$.  Then the restriction $\alpha \:C\sE\to E_{n+1}$
of~$\tilde{c}$ to~$C\sE$ satisfies~(i) and~(ii).  The converse is proved by
gluing the trivial map on~$CF$ to~$\alpha $. 
        \enddemo

In our application the inclusion $i\:F\to\sE$ is null homotopic, and it is
convenient to specify a null homotopy as follows. 

        \proclaim{\protag{5.5} {Lemma}}
 Let $F@>i>> \sE @>\pi >> B$ be a fibration, $\varphi \:F'\to F$ a continuous
map, and $H\:\zo\times F'\to\sE$ a null homotopy of~$i\circ \varphi $.  Then
a section $s\:F\to F'$ of~$\varphi $ determines a null homotopy~$H\circ s$
of~$i$.  Furthermore, the map $\pi \:(\sE,F)\to (B,b_0)$ is homotopy
equivalent to the based map 
  $$ \pi \vee (\pi \circ H\circ s)\:\sE\,\vee\,\Sigma F\longrightarrow
     B. \tag{5.6} $$
        \endproclaim

\flushpar
 Some definitions: $H$~is a null homotopy of~$i\circ f$ means $H_0=i\circ
\varphi $ and $H_1$~ maps to the basepoint of~$\sE$.  The section~$s$
satisfies~$\varphi \circ s=\id_F$.  The {\it wedge\/}~$X\vee Y$ of pointed
spaces~$X,Y$ is the union along the basepoints.  The map $\pi \circ H\circ
s\:\Sigma F\to B$ sends~$(t,f)\in \Sigma F$ to~$(\pi \circ H\mstrut _t\circ
s)(f)$.  The proof of \theprotag{5.5} {Lemma} is straightforward.
 
We apply these ideas first to the classifying space of~$G=SU_N,\,N\ge2$.  Now
`$E$'~denotes the specific cohomology theory described in~\S{1}. 

        \proclaim{\protag{5.7} {Proposition}}
 The classes~$\nu \in E^5(G)$ and~$\mu \in E^{6}(BG)$ are related by
transgression in the universal fibration $G @>i>> EG @>\pi >> BG$. 
        \endproclaim

        \demo{Proof}
 Let $\varphi =s\:G\to G$ be the identity map, $K$~a null homotopy
of~$\id_{EG}$, and $H=K\circ i\circ \varphi $.  The map~$\delta $
in~\thetag{5.3} is obtained by applying~$[\;\cdot \,,E_6]$ to the inclusion
$\Sigma F\to \sE\vee \Sigma F$.  (Recall that $[X,Y]$~is the set of homotopy
classes of maps~$X\to Y$.)  Set 
  $$ \psi =\pi \circ H\circ s\:\Sigma G\longrightarrow BG \tag{5.8} $$
to be the map in~\thetag{5.6}.  Notice that $j$ in~\thetag{5.3} is an
isomorphism for~$n=5$.  Therefore, we must prove that under $\psi
_E^*\:E^6(BG)\to E^6(\Sigma G)\cong E^5(G)$ we have $\psi _E^*(\mu )=\nu $.
In the commutative diagram
  $$ \xymatrix{ H^6(BG)\ar[r]^{\psi _H^*} \ar[d]_i &H^6(\Sigma G)\ar[d]^i
     &H^5(G)\ar[l]_-<<<{\cong } \ar[d]^i \\
      E^6(BG)\ar[r]^{\psi _E^*} \ar[d]_k &E^6(\Sigma G)\ar[d]^k
     &E^5(G)\ar[l]_-<<<{\cong }\ar[d]^k \\
      H^6(BG)\ar[r]^{\psi _H^*} &H^6(\Sigma G) &H^5(G)\ar[l]_-<<<{\cong }}
      $$
all groups are infinite cyclic, the maps~$i$ are isomorphisms, and $k\circ
i$~is multiplication by~2.  Parts~(ii) and~(iv) of \theprotag{1.9}
{Proposition} reduce us to showing~$\psi _H^*(c_3)=h_5$, which is a standard
transgression in the theory of characteristic classes in ordinary cohomology.
        \enddemo

We remark that the map~\thetag{5.8} classifies the bundle on~$\Sigma G$ whose
clutching function is $\id_G\:G\to G$; this is clear from its definition as
$\pi \circ H\circ s=\pi \circ K\circ i$.  It is also the first stage of the
Milnor construction of~$BG$.

        \demo{Proof of \theprotag{4.12} {Theorem}}
 From \theprotag{5.4} {Lemma} to produce the desired map $\alpha \:C\sE\to
E_{n+1}$ it suffices to show that $\nu $~and $\mu _1-\mu _2$~are related by
transgression in the fibration~\thetag{4.11}.  Consider the commutative
diagram
  $$ \xymatrix{
      &&E(G\times G)\ar[dr]^{\pi '}\ar[d]^q \\
      G\times G \ar[urr]^{H'} \ar[r]^\varphi & (G\times G)\!\bigm/\!G
     \ar[r]^-i\ar[d]_\chi  & E(G\times G)\!\bigm/\! G \ar[r]^-\pi
     \ar@{=}[d] & E(G\times G)\!\bigm/\! G\times G\ar@{=}[d]\\
      &G\ar[ul]^s \ar[r]^i &\sE \ar[r]^\pi &BG\times BG}
      \tag{5.9} $$
In this diagram $\varphi $~and $q$~are quotient maps; $s(g)=(g,e)$, where
$e\in G$~ is the identity element; the diffeomorphism~$\chi $ is given as
$\chi (g_1,g_2)=g_1g_2\inv $, which is well-defined on left cosets of~$G$;
and $H'$~is the null homotopy obtained from a null homotopy~$K\times K$
of~$EG\times EG$ (see the proof of \theprotag{5.7} {Proposition} above).
Then $H=q\circ H'$~is a null homotopy of~$i\circ \varphi $.  It follows that
$\pi \circ H\circ s\:\Sigma G\to BG\times BG$ in~\thetag{5.6} is~$\psi \times
*$, where $\psi $~is the map defined in~\thetag{5.8} and $*$~is the constant
map to the basepoint.  Note that $\sE\simeq BG$ and $\pi $~is homotopy
equivalent to the diagonal $\Delta \:BG\to BG\times BG$.  From the proof of
\theprotag{5.7} {Proposition} we see that $\mu _1-\mu _2\in E^6(BG\times BG)$
pulls back to $\nu \in E^5(G)\cong E^6(\Sigma G)\subset E^6(\sE\vee \Sigma
G)$ under
  $$ \xymatrix{\sE\vee\Sigma G \ar[r]^-{\simeq} &BG\vee\Sigma G
     \ar[rr]^{\Delta \vee(\psi \times *)} &&BG\times BG},  $$
and this completes the proof.
        \enddemo

As a preliminary to proving \theprotag{4.13} {Theorem} we discuss some
generalities about differential cohomology~\cite{HS,\S4}.  Let
$\bigl(C^\bullet(X;\RR),\delta  \bigr)$ denote the singular cochain complex
of a space~$X$.  In the differential theory we need to fix cocycles~$\iota
_n\in C^n(E_n;\RR)$ which represent the map $\frac 12k\:E_n\to H\RR_n$;
see~\thetag{1.4}.  These cocycles must satisfy the compatibility condition
$s^*\iota _{n+1}\!\bigm /\!Z_{\cir}=\iota _n$, where $s$~is the composition
$\cir\times E_n\to \Sigma E_n\to E_{n+1}$ and $Z_{\cir}$~is a fixed
fundamental cycle, paired with~$s^*\iota _{n+1}$ by slant product. 

        \definition{\protag{5.10} {Definition}}
 A {\it differential $E$-function of degree~$n$\/} on a smooth manifold~$M$
is a triple~$\cb=(\beta ,h,\omega )$ consisting of a continuous map $\beta \:M\to E_n$, a
{\it closed\/} differential form~$\omega $ of degree~$n$, and a cochain~$h\in
C^{n-1}(M;\RR)$ which satisfies
  $$ \delta h=\omega -\beta ^*\iota _n. \tag{5.11} $$
        \enddefinition

\flushpar
 In~\thetag{5.11} the differential form~$\omega $---termed the curvature
of~$\cb$---is regarded as a singular cocycle by integration.  A differential
$E$-function was termed an `$\cE$-cocycle' in~\S{4}.  A homotopy (or
morphism) is a differential $E$-function of degree~$n$ on~$\zo\times M$ whose
curvature is pulled back from~$M$ under projection.\footnote{In the language
of~\cite{HS} the condition on the curvature is captured by a filtration on
the space of differential functions: $\filt_0(E;\iota )^M$~is a category
whose set of isomorphism classes~$\pi _0\filt_0(E;\iota )^M$ is~$\cE^n(M)$.}
The set of equivalence classes under the induced equivalence relation is the
differential $E$-cohomology~$\cE^n(M)$.  Note that~$\omega $
in~$\cb=(\beta ,h,\omega )$ is an invariant of the class of~$\cb$ in~$\cE^n(M)$.
Analogous definitions apply to any generalized cohomology theory~$E$.  The
fundamental cocycles~$\iota _n$ then have coefficients in the vector
space~$E^0(\point;\RR)$.
 
To discuss transgression we also need a notion of `$\cE$-cochain'. 

        \definition{\protag{5.12} {Definition}}
 A {\it coned differential $E$-function of degree~$n$\/} on a smooth
manifold~$M$ is a triple~$\ca=(\alpha ,k,\eta )$ consisting of a continuous
map $\alpha \:CM\to E_{n+1}$, a differential form~$\eta \in \Omega ^n(M)$,
and a cochain~$k\in C^{n-1}(M;\RR)$.  It {\it trivializes\/} the differential
$E$-function $\cb=(\beta ,h,\omega )$ of degree~$n+1$, where
  $$ \aligned
      \beta &=\alpha \res M \\
      \omega &=d\eta \\
      h&=\eta -\alpha ^*\iota _{n+1}\!\bigm/\! Z\mstrut _{\zo} -\delta
     k.\endaligned \tag{5.13} $$
        \enddefinition

\flushpar
 In~\S{4} we termed~$\cb$ the `$\cE$-differential' of~$\ca$. The slant
product in~\thetag{5.13} is computed after pullback by the collapse
$\zo\times M\to CM$, and the cycle~$Z_{\zo}$ pushes to~$Z_{\cir}$ under the
collapse $\zo\to\cir$.  A homotopy is a coned differential
$E$-function~$(\Alpha ,K,\Xi )$ of degree~$n$ on~$\zo\times M$ such that the
restriction of~$\Alpha$ to $\zo\times M\subset \zo\times CM$ is constant,
$\Xi $~is pulled back from~$M$ under projection, and $H\in C^n\bigl(\zo\times
M;\RR \bigr)$ defined as (see~\thetag{5.13})
  $$ H=\Xi -\Alpha ^*\iota _{n+1}\!\bigm/\!Z\mstrut _{\zo} -\delta K
     \tag{5.14} $$
is also pulled back from~$M$ under projection.  The set of equivalence
classes of coned differential $E$-functions which trivialize a fixed
differential $E$-function of degree~$n+1$ is a torsor for~$\cE^n(M)$.
 
Integration of differential functions is defined in~\cite{HS,\S4.10}.  Our
definition~\thetag{4.16} of the WZW~factor requires that integration be
extended to {\it coned\/} differential functions as well.  The main step is
simply extending the orientation map, called~`$\mu $' in~\cite{HS,\S4.10}, to
the path spaces in the spectrum.\footnote{A based map $CX\to E_{n+1}$ is by
adjunction a map $X\to PE_{n+1}$ to the space of paths beginning at the
basepoint of~$E_{n+1}$.}

        \demo{Proof of \theprotag{4.13} {Theorem}}
 We must construct the coned differential $E$-function $\ca=(\alpha ,k,\eE )$
on~$\sE$.  The function~$\alpha \:C\sE\to E_6$ is the topological
transgressing `$E$-cochain' in \theprotag{4.12} {Theorem}.  We take $k=0$, but
discuss other possibilities following the proof.  Our main task is to
construct a transgressing form~$\eE \in \Omega ^5(\sE)$.
 
Recall first the Chern-Simons~\cite{CS} transgressing form in the universal
bundle $G\to EG\to BG$.  Denote by $\langle \cdot ,\cdot ,\cdot \rangle$ the
symmetric trilinear form~\thetag{3.3} on $\frak{g}=\Lie(G)$.  Let $\Theta
=\Theta ^{\text{univ}}\in \Omega ^1_{EG}(\frak{g})$ be the universal
connection on~$EG$ and $\Omega \in \Omega ^2_{EG}(\frak{g})$ its curvature.
In this notation the 6-form in~\thetag{1.14}, lifted to~$EG$, is
  $$ \pi ^*\wm = \langle \Omega \wedge \Omega \wedge \Omega
     \rangle.  $$
Then the Chern-Simons 5-form
  $$ \eta =\langle \Theta \wedge \Omega \wedge \Omega \rangle -\frac
     14\langle \Theta \wedge \Omega \wedge [\Theta \wedge \Theta ] \rangle +
     \frac{1}{40}\langle \Theta \wedge [\Theta \wedge \Theta ]\wedge [\Theta
     \wedge \Theta ] \rangle \tag{5.15} $$
satisfies the transgression conditions
  $$ \align
      &\text{(i)\ the restriction of~$\eta$ to a fiber equals~$\wn$; and}
     \\
      &\text{(ii)\ $d\eta=\pi ^*\wm$.}
      \endalign $$
In other words, $\eta $~is a transgressing 5-form for the curvatures in
\theprotag{5.7} {Proposition}.  

We want now to construct a transgressing 5-form~$\eE$ in the fiber bundle
$\sE @>\pi >> BG\times BG$ with fiber $(\GG)/G\cong G$. Recall that
$\sE=(EG\times EG)/G$.  We write the pullback of~$\eE$ to~$EG\times EG$, and
express it in terms of the universal connection forms~$\Theta _1,\Theta _2$
and universal curvature forms~$\Omega _1,\Omega _2$ on~$EG\times EG$, but
where these forms are regarded as $\frak{g}$-valued by identifying
$\frak{g}_1\cong \frak{g}_2\cong \frak{g}$.  Write~$\eta _1$
for~\thetag{5.15} with $\Theta _1,\Omega _1$ replacing~$\Theta ,\Omega $ and
similarly for~$\eta _2$.  Then
  $$ \eE=\eta _1 - \eta _2 + d\tau , \tag{5.16} $$
where 
  $$ \multline  
      \tau = \frac 14 \langle \Theta _1\wedge \Theta _2\wedge [\Theta _1\wedge
     \Theta _1] \rangle + \frac 14 \langle \Theta _1\wedge \Theta _2\wedge
     [\Theta _2\wedge \Theta _2] \rangle - \frac 14 \langle \Theta _1\wedge
     \Theta _2\wedge [\Theta _1\wedge \Theta _2] \rangle \\ 
      - \langle \Theta
     _1\wedge \Theta _2\wedge \Omega _1 \rangle - \langle \Theta _1\wedge
     \Theta _2\wedge \Omega _2 \rangle.\endmultline  $$
The form~$\eE$ satisfies
  $$ \align
      &\text{(i)\ $\eE$~is basic for the diagonal $G$-action on~$EG\times
      EG$;} \\
      &\text{(ii)\ the restriction of~$\eE$ to a fiber of $\pi \:\sE\to
     BG\times BG$ is cohomologous to~$\wn$; and} \\
      & \text{(iii)\ $d\eE=\pi ^*(\omega _{\cm_1} - \omega _{\cm_2})$.}
      \endalign $$
We leave the reader the exhausting task of checking these properties.  It
helps to observe that the pullback of~$\wn$ under the diffeomorphism $\chi
\:(G\times G)/G\to G$ in \thetag{5.9} is the restriction of~$\eE$ to a
fiber. 
        \enddemo

The choice of~$k=0$ in $\ca=(\alpha ,k,\eE)$ fixes the differential function
representing $\cm_1-\cm_2$ to be the triple $\cb=(\alpha
\res\sE\;,\;\sE-\alpha ^*\iota _6\!\bigm/\! Z_{\zo}\;,\;\omega
_{\cm_1}-\omega _{\cm_2})$; see~\thetag{5.13}.  Any other choice of~$k\in
C^4(\sE;\RR)$ such that $\ca$~trivializes $\cb$ must satisfy~$\delta k=0$, so
determines a class $[k]\in H^4(\sE;\RR)$.  Since equivalence classes of
trivializations of~$\cb$ is a torsor for~$\cE^5(\sE)$---see the remark
following~\thetag{5.14}---only its image in $H^4(\sE;\RZ)$ is relevant.  In
other words, equivalence classes of possible~$\ca$ are parametrized by
elements of $H^4(\sE;\RZ)\cong H^4(BG;\RZ)\cong \RZ$.  Write
$[k]=\frac{\theta }{N_c}c_2$, where $c_2\in H^4(BG)$ is the generator and
$\theta \in \RR$ is determined modulo~$N_c\ZZ$.  Recall from the end of~\S{4}
that the $(\GG)$-bundle $Q\to X$ in~\thetag{4.15} has a reduction to a
$G$-bundle $\bQ\to X$.  Then a nonzero cohomology class~$[k]$ multiplies the
WZW~factor~\thetag{4.15} by 
  $$ \exp\left( \tpi \,\theta \int_{X}c_2(\bQ) \right).   $$
This matches the topological term~\thetag{3.9} in gauged~QCD with $\theta
=\theta _1+\theta _2$. 
 
To conclude we sketch an argument proving that the gauged WZW factor~$\tW_X$
in \theprotag{4.14} {Definition} is well-defined.  It is a function of a
quartet~$q=(Q,\Theta ,\gamma ,\phi )$ which is an object in a groupoid~$\sG$.
More precisely, there is a hermitian line bundle $\sL\to\sG$ and $\tW_X$~is
meant to be a section of the line bundle.  This means that there is a complex
line~$\sL_q$ attached to each object~$q$, a linear isomorphism $\epsilon
\:\sL_q\to\sL_{q'}$ attached to each morphism $q\to q'$, and
  $$ \epsilon \bigl(\tW_X(q) \bigr) = \tW_X(q'). \tag{5.17} $$
It is this last point which we must check.  First, we recall that there is a
groupoid~$\sG_2$ whose objects are triples~$(Q,\Theta ,\gamma )$ of a
principal $(\GG)$-bundle $Q\to X$, a connection~$\Theta $, and a classifying
map $\gamma \:Q\to E(\GG)$ for the connection.  A morphism $(Q,\Theta ,\gamma
)\to (Q',\Theta ',\gamma ')$ is an equivalence class of
quintuples~$(P,\Lambda ,\Gamma ,\varphi _0,\varphi _1)$ consisting of a
$(\GG)$-bundle $P\to \zo\times X$; a connection~$\Lambda $ on~$P$ whose
curvature~$\Omega $ satisfies~$\iota (\partial /\partial t)\Omega =0$, where
$t$~is the coordinate on~$\zo$; a classifying map~$\Gamma \:P\to E(\GG)$ for
the connection~$\Lambda $; an isomorphism~$\varphi _0$ of~$(Q,\Theta ,\gamma
)$ with the restriction of~$(P,\Lambda ,\Gamma )$ to~$\{0\}\times X$; and an
isomorphism~$\varphi _1$ of~$(Q',\Theta ',\gamma ')$ with the restriction
of~$(P,\Lambda ,\Gamma )$ to~$\{1\}\times X$.  Quintuples~$(P,\Lambda ,\Gamma
,\varphi _0,\varphi _1)$ and~$(P',\Lambda ',\Gamma ',\varphi _0',\varphi
_1')$ are identified if there is an isomorphism $P\to P'$ which preserves the
connections and isomorphisms~$\varphi _i,\varphi '_i$ and under which $\Gamma
$~and $\Gamma '$~are homotopic.  An object~$(Q,\Theta ,\gamma ,\phi )$ in the
groupoid~$\sG$ includes the section~$\phi $ of the associated bundle $G_Q\to
X$, and likewise a morphism~$(P,\Lambda ,\Gamma ,\Phi ,\varphi _0,\varphi
_1)$ includes a section~$\Phi $ of $G_P\to\zo\times X$ which satisfies
$\nabla _{\partial /\partial t}\Phi =0$, i.e., $\Phi $~is flat along
trajectories of~$\partial /\partial t$.  Given such a morphism we compute
  $$ \exp\left( \tpi\int_{\zo\times X}N_c\,\Phi ^*\tG^*\ca \right),
     \tag{5.18} $$
($\tG\:G_P\to\sE$ is the induced classifying map) and apply an appropriate
version of Stokes' theorem.  The `$\cE$-differential' of~\thetag{5.18} is
computed in terms of differential forms as
  $$ \exp\left( \tpi\int_{\zo\times X}N_c\,\Phi ^*\tG^*\eE
     \right).  $$
This equals~1 because of the ``constancy'' of~$\Phi $ and~$\Lambda $ in the
$\partial /\partial t$~direction.  The integral over~$\zo\times X$ of the
`$\cE$-differential' of the integrand in~\thetag{5.18} is a linear
isomorphism $\epsilon \:\sL_{(Q,\Theta ,\gamma )}\to\sL_{(Q',\Theta ',\gamma
')}$, and the integral over the boundary of~$\zo\times X$ is the ratio
of~$\tW_X(Q',\Theta ',\gamma ')$ to $\tW_X(Q,\Theta ,\gamma )$.  Stokes'
theorem then implies~\thetag{5.17} immediately.

\widestnumber\key{SSSSSSSS}   

\enddocument